\documentclass[notitlepage,twocolumn,prx,nobalancelastpage,amssymb,superscriptaddress,showpacs,longbibliography]{revtex4-1}

\usepackage{Symbols}
\usepackage{Shortcuts}
\usepackage{Text}

\usepackage{stmaryrd}
\usepackage{epsfig}
\usepackage{subfigure}
\usepackage[colorlinks=true,citecolor=blue,linkcolor=blue]{hyperref}
\usepackage{verbatim}
\usepackage{xcolor}

\def \vc{\boldsymbol{c}}
\newcommand{\addMR}[1]{\textcolor{red}{#1}}

\begin{document}
\title{Universal transport in periodically driven systems without long-lived quasiparticles}
%Quantization of current and thermalization in topological pumps with SYK interactions

\author{Iliya Esin}
\affiliation{Department of Physics, California Institute of Technology, Pasadena, CA 91125, USA}
\author{Clemens Kuhlenkamp}
\affiliation{Department of Physics, Technical University of Munich, 85748 Garching, Germany}
\affiliation{Munich Center for Quantum Science and Technology (MCQST), Schellingstr. 4, D-80799 M\"unchen, Germany}
\affiliation{Institute for Quantum Electronics, ETH Z\"urich, CH-8093 Z\"urich, Switzerland}
\author{Gil Refael}
\affiliation{Department of Physics, California Institute of Technology, Pasadena, CA 91125, USA}
\affiliation{Walter Burke Institute for Theoretical Physics, California Institute of Technology, Pasadena, CA 91125, USA}
\author{Erez Berg}
\affiliation{\mbox{Department of Condensed Matter Physics, Weizmann Institute of Science, Rehovot, 76100, Israel}}
\author{Mark S. Rudner}
\affiliation{Department of Physics, University of Washington, Seattle, WA 98195-1560, USA}
%\affiliation{\mbox{Center for Quantum Devices and Niels Bohr International Academy,}
%\mbox{Niels Bohr Institute, University of Copenhagen, 2100 Copenhagen, Denmark}}
\author{Netanel H. Lindner}
\affiliation{Physics Department, Technion, 3200003 Haifa, Israel}

\date{\today}

\begin{abstract}
An intriguing regime of universal charge transport at high entropy density has been proposed for periodically driven interacting one-dimensional systems with Bloch bands separated by a large single-particle band gap. For weak interactions, a simple picture based on well-defined Floquet quasiparticles suggests that the system should host a quasisteady state current that depends only on the populations of the system's Floquet-Bloch bands and their associated quasienergy winding numbers. Here we show that such topological transport persists into the strongly interacting regime where the single-particle lifetime becomes shorter than the drive period. Analytically, we show that the value of the current is insensitive to interaction-induced band renormalizations and lifetime broadening when certain conditions are met by the system's non-equilibrium distribution function. We show that these conditions correspond to a quasisteady state. We support these predictions through numerical simulation of a system of strongly interacting fermions in a periodically-modulated chain of Sachdev-Ye-Kitaev dots. Our work establishes universal transport at high entropy density as a robust far from equilibrium topological phenomenon, which can be readily realized with cold atoms in optical lattices.
% We investigate the dynamics of a periodically driven strongly interacting system described by a model of a multi-orbital non-adiabatic topological Thouless pump. Fermions occupying different orbitals can locally interact through random interactions, realizing a copy of a charged Sachdev-Ye-Kitaev (SYK) dot at each site. While each single SYK dot is maximally chaotic and is known as an effective thermalizer, a periodically driven chain of such dots exhibits a regime in which the thermalization is exponentially slow, giving rise to a quasi-steady state. In this quasi-steady state, fermions uniformly occupy single-particle states in one of the chiral Floquet bands, resulting in a quantized charge transfer per cycle of the drive. The transport in the quasi-steady state obtains a universal value, even in a partially filled system, in contrast to non-interacting adiabatic Thouless pumps. Our work opens a new avenue in the studies of novel dynamical phases in strongly coupled materials, such as high-temperature superconductors and moir\'e materials under non-equilibrium conditions.
\end{abstract}

\maketitle

\section{Introduction}

% \subsection{Background for topological pumps}

%\addMR{\bf Make sure we cite Floquet review by Eckardt: Rev. Mod. Phys. 89, 011004; also Nigel Cooper (check our list of cited reviews in our Nature Reviews Physics intro for others)}
The interplay of topology and far from equilibrium dynamics became an important arena of research in recent years~\cite{Oka2009,Kitagawa2010,Lindner2011,Dalibard2011,Gomez-Leon2013,Huse2013,Rudner2013,Dehghani2014,FoaTorres2014,Usaj2014,Chandran2014,Goldman2014,Grushin2014,Perez-Piskunow2015,DalLago2015,Nathan2015,Khemani2016,VonKeyserlingk2016,VonKeyserlingk2016a,Else2016,Po2016,Potter2016,Eckardt2017,Roy2017,Roy2017a,Harper2017,Gong2018,Esin2018,Ozawa2019,Cooper2019,Rudner2020a,Harper2020,Zhang2021}. In equilibrium, the field of topology has substantially influenced the modern understanding of electronic systems, contributing to the introduction of fundamental concepts such as topological robustness of quantum states, topological degeneracies of ground states, and non-abelian anyonic statistics~\cite{Arovas1985,Altland1997,Kitaev2006,Stern2008,Nayak2008,Kitaev2009,Hasan2010,Qi2011,Fidkowski2011,Ren2016,Chiu2016,Wen2017}. Extension of these concepts to non-equilibrium systems provides the means for dynamical control of topological properties and design of topological phases ``on-demand''~\cite{Aidelsburger2011,Rechtsman2013,Aidelsburger2014,Jotzu2014,Fleury2016,Basov2017,Oka2019,McIver2019,Shan2021,Mukherjee2021}. 
%\addMR{\bf Aidelsburger Measuring Chern number: Nature Physics volume 11, pages162–166 (2015)]}
Recently, periodic drives were employed to induce exotic phases of matter without equilibrium analogs~\cite{Else2016a,Titum2016,Peng2016,Zhang2017,Choi2017,Maczewsky2017, Mukherjee2017a,Nathan2019, wintersperger2020,Afzal2020,Esin2020,Esin2021}. %\addMR{\bf [Somewhere add reference to Rechtsman topological soliton PhysRevX.11.041057. Also be sure to include Foa Torres, Usaj, etc., Bukov, Fleckenstein + Bukov, Fiete PhysRevX.9.021037? Polkovnikov; more preth]}

\begin{figure}[h!]
  \centering
  \includegraphics[width=8.6cm]{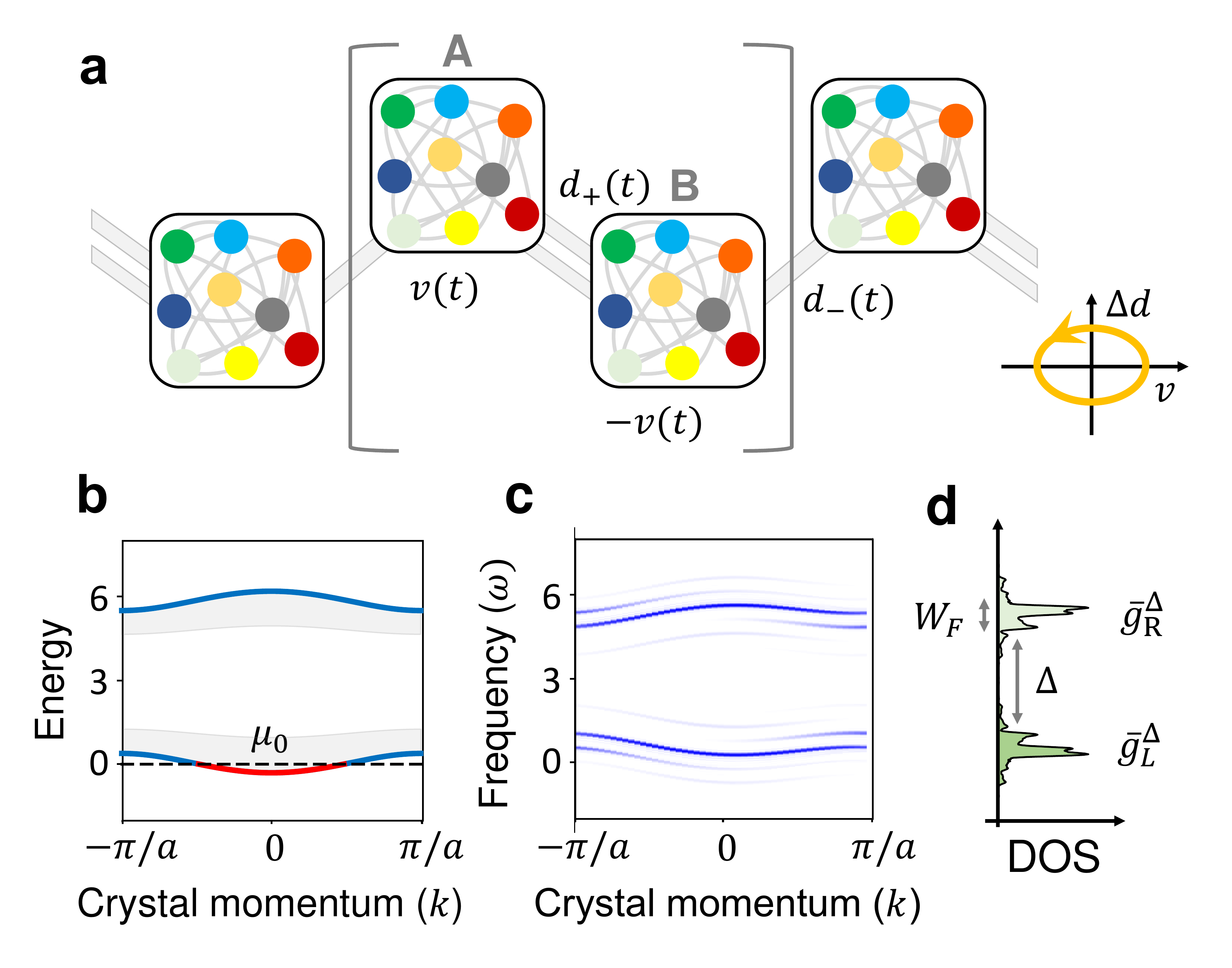}\\
  \caption{\tb{The model and the single-particle spectrum.} \tb{a} Illustration of the system. A chain of atoms with time-modulated hopping and staggered potential, following the driving protocol realizing an adiabatic pump, as shown in the right bottom corner of panel \tb{a} (we denote $\D d(t)=d_+(0,t)-d_-(0,t)$). In the numerical simulation, each site is an SYK dot with $N\to \infty$ orbitals that interact through random interactions [see Eq.~\eqref{eq:SYKHamiltonian}].
 \tb{b} The single-particle spectrum of the system at $t<0$, before the drive is switched on. The dashed line represents the Fermi level and the colored section represents initially occupied states, constituting a quarter of all the states in the system. The shaded gray areas show the instantaneous levels of the driven system in one period. The band structure is calculated for $J_0=1$, $J_1=0.85$, $v_0=2.55$, and $\W=0.5$ [see Eq.~\eqref{eq:HamiltonianRM} for the definition of the single-particle Hamiltonian].
\tb{c} The period averaged spectral function of the non-interacting system, $\bar g^\D(k;\w)$.
Energy and frequency in panels \tb{b} and \tb{c} are plotted in units of $J_0 = 1$.
\tb{d} The period-averaged non-interacting density of states (DOS). \label{fig:System}}
\end{figure}

An important paradigmatic model of an intrinsically non-equilibrium topological system is the topological pump. 
The topological pump, originally introduced by Thouless~\cite{Thouless1983}, describes a one-dimensional atomic chain with an adiabatically slowly and periodically in time modulated potential. Such a system when tuned to its topological phase, supports a robust quantized transport~\cite{Niu1984,Niu1990,Brouwer1998,Switkes1999,Altshuler1999,Chern2007,Xiao2010,Meidan2011,Wang2013}. The precise quantization and robustness of adiabatic pumps to external perturbations makes them important candidates for applications in quantum metrology~\cite{Keller1999,Piquemal2004,Blumenthal2007,Giblin2012,Pekola2013} and processing of quantum information~\cite{Fletcher2013,Ubbelohde2014,Johnson2017}. % storage of charge [Ref:cite patent [Klaus Ensslin] Quantum wire CCD charge pump.]. 
Adiabatic pumps were recently experimentally realized in photonic systems and cold atoms~\cite{Kraus2012,Schweizer2016,Lohse2016,Nakajima2016,Zilberberg2018,Lohse2018,Cerjan2020,Nakajima2021,Jurgensen2021,Minguzzi2021}.

The realization of topological %adiabatic 
pumps in metallic interacting systems is challenging %\addMR{\bf [MR: not so bad for gapped systems]} 
due to an interplay of inter-particle interactions and the non-adiabatic evolution stimulated by the periodic drive. %\addMR{\bf [MR: also, meaning of adiabaticity for metal?]} 
Such an interplay often results in an extensive generation of entropy and incessant heating up of the system to a featureless, high-entropy state~\cite{Lazarides2014,DAlessio2014,Ponte2015}. The heating can be significantly slowed down in the high driving frequency regime or under special conditions, giving rise to a long-lived prethermal state \cite{Bukov2015,Eckardt2015,Abanin2015,Bukov2016,Kuwahara2016,Else2017,Abanin2017,Reitter2017,Mori2018,Howell2019,Vogl2019,Kuhlenkamp2020,Fleckenstein2021}. %  which is often detrimental for the coherent quantum state. In a generic quantum system, such an interplay
Recently it was shown that %in the weakly interacting limit and under certain conditions the system forms 
 a slowly driven topological pump in the weakly interacting limit can form a quasi steady state~\cite{Lindner2017,Gulden2020,Gawatz2021}.
In this limit, the quasi-steady state can be understood heuristically on the level of free dynamics and weak scattering of particles in well-defined Floquet-Bloch bands. %The quasi steady-state results from the quasiparticle scattering between the weakly interacting Floquet-Bloch states. 
Notably, the quasi steady state hosts a universal current that depends only on the populations of the system’s Floquet-Bloch bands and their associated quasienergy winding numbers. %\addMR{\bf [MR: somehow make it clear that this is a heuristic picture based on weakly/non-interacting particles in well-defined Floquet-Bloch bands.]}

% Floquet-Bloch states and due to scattering of the quasi-particles between the Floquet-Bloch states. The intuition behind this result follows from quasiparticle scattering which 
% Intuitively, this result is  be understood as the quasiparticles 
% The intuitive understanding of the steady state  behind the quasi state state comes from 

% the unique structure of the Floquet-Bloch states 

% the unique structure of the Floquet states in the narrow bandwidth, weakly interacting adiabatic pumps, gives rise to a scale separation, leading to an exponentially long-lived quasi steady state \cite{Lindner2017,Gulden2020}. 
% or weak interactions, a simple picture based on well-defined Floquet quasi-particles suggests that the system should host a quasisteady state current that depends only on thepopulations of the system’s Floquet-Bloch bands and their associated quasienergy winding numbers.

% Slow thermalization of the quasi steady state to a maximal entropy state was demonstrated using a perturbation theory and the exact evolution of a small system in time. 

%It was also conjectured that the quasi steady state is a non-equilibrium state where one of the bands has an infinite temperature with a fractional occupation while the other band is unoccupied.

Here we show that the quasi-steady state persists into the strongly interacting regime where the single-particle scattering lifetime is shorter than the drive period. 
Furthermore, in this regime, the current exhibits a similar universal value as in the weakly interacting case, despite the absence of long-lived single-particle Floquet states. %despite strong renormalization of the  Floquet-Bloch bands \addMR{\bf [MR: despite the absence of long-lived single-particle Floquet states (or something similar)?]}.  
We support our analytical predictions through numerical simulations of a system of strongly interacting fermions in a periodically-modulated chain of Sachdev-Ye-Kitaev (SYK) \cite{Sachdev1993,Kitaev2015,Maldacena2016} dots. Our work demonstrates a new approach for numerically exact simulations of %long
driven, strongly interacting chains of many sites, by specializing to SYK-type interactions. This method outperforms the conventional exact diagonalization methods that can be applied to significantly smaller systems. In turn, approximate methods such as Hilbert space decimation (i.e., the time-dependent density matrix renormalization group~\cite{White2004,Schollwock2005}), can not be applied here, because thermalization dynamics generates long-ranged correlations.

\section{Definition of the problem}
%To test the quasi steady state, we introduce a model of an interacting Thouless pump. 
%We begin the discussion with a definition of the specific model that we used to test analytically and numerically the quasi steady state regime. 
In this work, we consider a one-dimensional bipartite chain of $L$ unit cells with periodic boundary conditions, %\addMR{\bf [MR: PBC?]} 
hosting $N$ flavors of otherwise spinless fermions (see Fig.~\ref{fig:System}a).
We label the two sublattices by $A$ and $B$, %\addMR{\bf [MR: add $A$ and $B$ labels on figure]}
and denote the lattice constant by $a$. 
For %clarity of the 
simplicity of notation, throughout we set $\hb=k_B=e=1$.

At times $t\ge 0$, the evolution of the system is described by the time-periodic Hamiltonian  $\hat\cH(t)=\sum_{k\a} \hat \vc\dg_{k\a}H_{0}(k,t)\hat \vc_{k\a}+\hat\cH_{\rm int}$, where
% \Eq{
% H_{0}(k,t)=\mat{\m_0+v(t) & d_1(t)+d_2(t)e^{-i k a} \\ d_1(t)+d_2(t)e^{i k a} & \m_0 -v(t) },
% \label{eq:HamiltonianRM}
% }
$H_0(k,t)$ denotes the time-periodic single particle Bloch Hamiltonian and $\hat\cH_{\rm int}$ denotes the electron-electron interactions. 
Here, $\hat\vc_{k\a}\dg=(\hat c_{A,k\a}\dg,\hat c_{B,k\a}\dg)$; $\hat c_{s,k\a}\dg=\frac{1}{\sqrt{L}}\sum_{j\in j_s} e^{i k a j/2}\hat c_{j\a}\dg$, where $\hat c^\dagger_{j\a}$ creates a fermion at a position $j$ with flavor index $\a$; $j_s$ includes all the odd (even) sites for $s=A(B)$. %chain index \addMR{\bf [MR: ``chain'' or ``flavor'' index? also, some dependence on $s$ or a prime on the sum in the defintion of $c^\dagger_{s,k\alpha}$ might be helpful.]} $\a$. The $j$-sum in the definition of $\hat c_{s,k\a}\dg$ is over odd (even) sites for $s=A\,(B)$. 
The single-particle Hamiltonian describes the Rice-Mele model~\cite{Rice1982} with time-periodically modulated parameters: 
\Eq{
H_{0}(k,t)=\mat{v(t)-\m_0 & d(k,t) \\ d\as(k,t) &  -v(t)-\m_0 }.
\label{eq:HamiltonianRM}
}
Here, $d(k,t)=d_+(k,t)+d_-(k,t)$ and $d_{\pm}(k,t)=e^{\mp i k a/2}[J_0 \pm J_1 \sin(\W t)]$  %\addMR{\bf [MR: %too many different numbers on indices. Let's clean up notation. $d_\pm$? 
%Technically, with the chain as drawn, the phase factors should probably be equally split between the $d_1$ and $d_2$ terms, though it won't change anything for us.]} 
and $v(t)=v_0\cos(\W t)$, where $\W$ is the driving frequency and  $v_0$, $J_{0}$, $J_1$ are constants. %\addMR{\bf [Can we change $s_0$ and $s_1$ to $J_0$ and $J_1$?]}.
%The hopping coefficients $d_{1,2}(t)$ and the staggered potential $v(t)$ are periodically modulated in time, realizing the topological Rice-Mele model \cite{Rice1982}. In particular, we consider $d_{1,2}(t)=s_0 \pm s_1 \sin(\W t)$ and $v(t)=v_0\cos(\W t)$, where $\W$ is the driving frequency and  $v_0$, $s_{0}$, $s_1$ are constants. 
The chemical potential 
$\m_0$ sets the average density of the fermions in the chain % \cBlue{by shifting the total energy relative to the chemical potential fixed at $0$
(see Fig.~\ref{fig:System}b). %\addMR{\bf [MR: it's very confusing to call it $\mu$, then. Either we should put a $-$ and call it $\mu$, or change the name to $\varepsilon_0$ or something like that.]}.
For $t<0$, the system is assumed to be in an equilibrium state with respect to the Hamiltonian $\hat{\mathcal{H}}(t = 0)$, at inverse temperature $\be_0$. % described by a constant in time Hamiltonian, that we set to $\hat \cH(t<0)=\hat \cH(t=0)$, to ensure a smooth evolution at $t=0$.

The interparticle interactions are described by the Hamiltonian %\addMR{\bf [MR: ordering of operators?]}
\Eq{
\hat\cH_{\rm int}=\sum_{jj'}\sum_{\a\be\g\dl}U_{\a\be\g\dl}(j,j')\hat c_{j\a}\dg\hat c_{j\be}\hat c_{j'\g}\dg\hat c_{j'\dl}+{\rm h.c.}
\label{eq:SYKHamiltonian}
}
In our analytical study, we assume a single flavor, %chain \addMR{\bf [MR: flavor?]}
$\a=1$, and generic short-ranged interactions of characteristic strength $U_{\a\be\g\dl}(j,j')=U \chi(|j-j'|)$, where $\chi(|j-j'|)$ is a rapidly decaying dimensionless function of its argument, with $\chi(1)=1$. (Note that for a single species of fermions, the on-site interaction terms, $j = j'$, do not contribute.) %\addMR{(we leave the precise form unspecified)}. 
In the numerical study, we consider the limit of a large number of flavors, $N$, with $N\to\infty$, % $N$ chains, extending the chain index $\a$ to the range $[1,N]$ and taking the limit $N\to \infty$, 
see Fig.~\ref{fig:System}a for an illustration. Particles of different flavors %in different chains 
can locally interact through an SYK-type on-site interaction term,
%, described by 
% \Eq{
% \hat\cH_{\rm int}=\sum_{j,\a\be\g\dl}U_{\a\be\g\dl}\hat c_{j\a}\dg\hat c_{j\be}\dg\hat c_{j\g}\hat c_{j\dl}+{\rm h.c.}
% \label{eq:SYKHamiltonian}
% }
%Here, $\a,\be,\g,\dl=1,...,N$ and we take the limit $N\to \infty$. 
%Here,
where we consider random and constant in space interactions with $\overline {U_{\a\be\g\dl}(j,j')}=0$ and $\overline { U_{\a\be\g\dl}^2(j,j')}=\dl_{jj'}U^2/N^3$, such that the system preserves invariance to translations for every realization of disordered couplings~\cite{Chowdhury2018,Patel2018}. %We also include a weak random quadratic term, $\hat\cH_{\rm SYK-2}=\sum_{j,\a\be}K_{\a\be}\hat c_{j\a}\dg\hat c_{j\be}+{\rm h.c.}$, where $\overline{K_{\a\be}}=0$, $\overline{K_{\a\be}^2}=K^2/N$, which is essential to stabilize the numerics in the weakly interacting regime. 
% Our model  realizes a chain of coupled SYK dots\cite{Chowdhury2018} [add more refs], see Fig.~\ref{fig:System}a for an illustration.  

% In what follows, we study the system's dynamics through the time-evolved Green's functions \addMR{\bf [MR: do we need to include this technical text here, or can we just introduce each Green's function when it appears?]} providing information about expectations of the single-particle operators, such as particle densities and the current (see below). 
% The single-particle lesser Green's function \addMR{(with matrix structure in sublattice space)} is defined as $G^{<}_{ss'}(k;t,t')=\frac{i}{N}\sum_{\a=1}^N\overline{\av{\hat c\dg_{s,k\a}(t)\hat c_{s',k\a}(t')}}_{\ro_0}$,
% where the expectation value is calculated with respect to the initial state described by the density matrix $\ro_0$, and the bar denotes averaging over the random interaction strength. Similarly, the retarded function is given by 
% $G^{R}_{ss'}(k;t,t')=\frac{i}{N}\q(t-t')\sum_{\a=1}^N\overline{\av{\{\hat c\dg_{s,k\a}(t),\hat c_{s',k\a}(t')\}}}_{\ro_0}$, where ``$\bC{,}$'' denotes an anticommutator. The other components of the Keldysh Greens' function are defined in Appendix~\ref{sec:BareGreen}. Throughout, we omit the sublattice indices $s,s'$, assuming implicitly matrix products of the Green's functions, unless otherwise specified.  

\begin{figure}
  \centering
  \includegraphics[width=8.6cm]{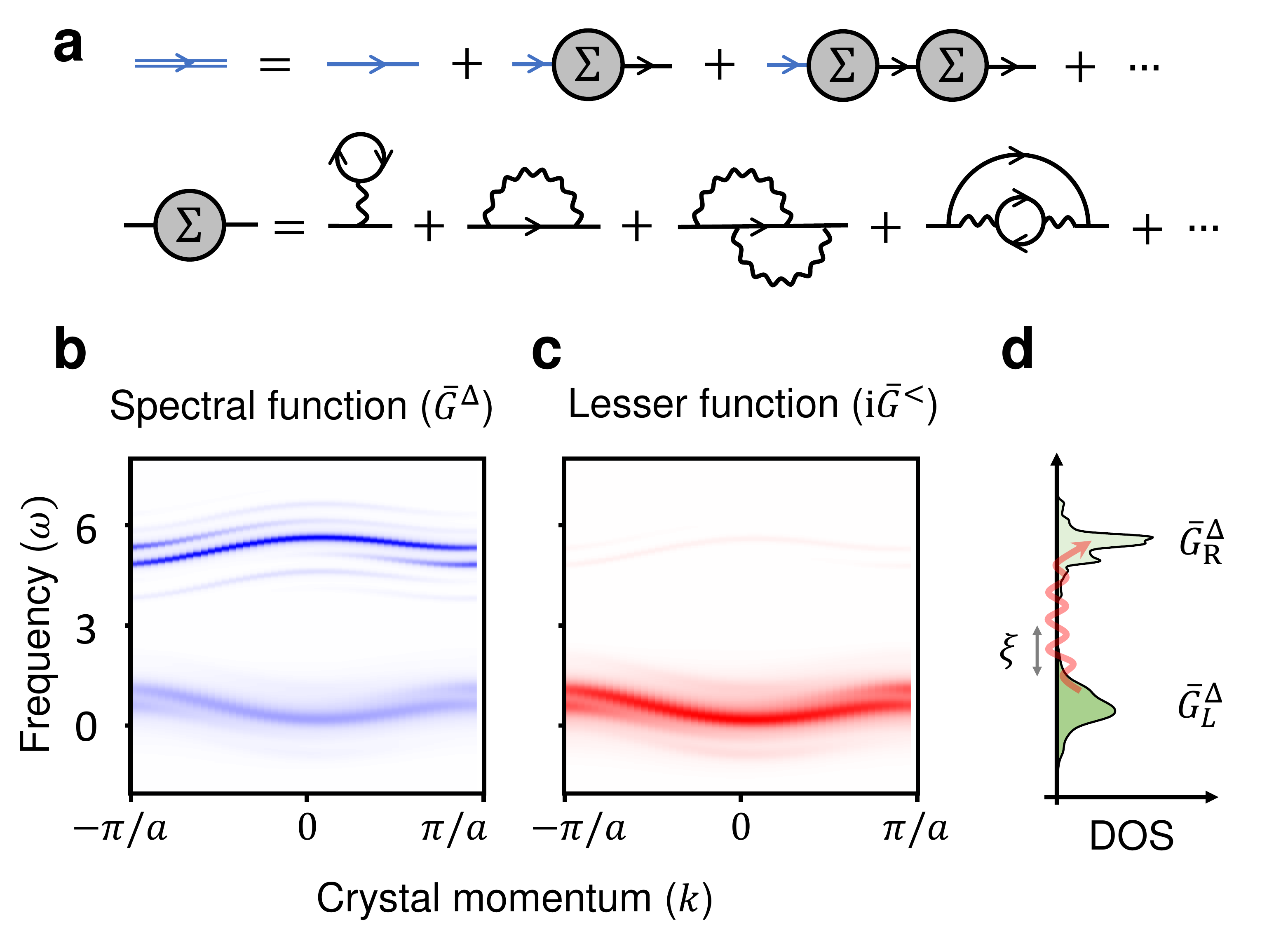}\\
  \caption{\tb{Renormalization of the spectrum by interactions.} \tb{a} First row: Dyson's expansion for the band resolved Green's function. The double and single blue lines indicate the band-resolved renormalized and bare functions respectively. Black lines denote the full bare Green's function.  Second row:  Diagrammatic expansion of the self-energy, used in the analytical model (shown up to order $U^2$). In the numerics, we considered SYK interactions in the infinite $N$ limit, in which all the diagrams with odd number of interaction vertices or with crossed lines are averaged to zero (i.e., only the last diagram in \tb{a} contributes). \tb{b} Period-averaged spectral function, $\bar G^\D(k;\w)$, renormalized by the SYK interactions with $U=1$.  \tb{c} Period-averaged lesser function, $i\bar G^<(k;\w)$, indicating occupation after 30 periods of the drive, for the same parameters as in \tb{b}. Note the weak, yet finite occupation of the upper band produced by the interband processes. \tb{d} The renormalized period-averaged single-particle DOS. In the presence of interaction, the non-interacting DOS appearing in Fig.~\ref{fig:System}d, broadens obtaining exponential tails $\x$. The broadening creates an overlap between the upper and lower bands, forming an interband heating channel (indicated by the wiggly arrow). Frequencies are plotted in units of $J_0=1$. \label{fig:Interacting}}
\end{figure}

\section{Non-interacting dynamics\label{sec:nonInter}}

Before studying the interacting model, we briefly summarize the dynamics of the non-interacting topological pump~\cite{Thouless1983,Shih1994,Xiao2010,Kitagawa2010,Lindner2017,Privitera2018}. We initialize the pump in an equilibrium state of $\cH(0)$, at inverse temperature $\be_0$ and a chemical potential $\mu_0$ that fixes the average density of particles at $n_0$, see Fig.~\ref{fig:System}b. The spectral function is initially periodic in $k$, following the spectrum of the static Hamiltonian, $\cH(0)$, as is demonstrated in Fig.~\ref{fig:System}b. 

After switching on the drive (i.e., for $t>0$), the dynamics of the system follows the time-dependent Hamiltonian $\cH(t)$. 
Shortly after the quench, the bands of high intensity in the spectral function develop a pronounced structure of sidebands, spaced by the drive frequency $\Omega$, and furthermore obtain nonvanishing net slopes~\cite{Kitagawa2010} as a function of $k$ %, \cBlue{and side bands shifted by $\W$} 
(see Fig.~\ref{fig:System}c and attached video). The peaks of the spectral function at each $k$ correspond to  quasienergies associated with the single-particle Floquet state solutions~\cite{Rudner2020,Rudner2020a} of  %This form of the spectral function follows \addMR{follows how? mention delta function peaks at quasienergies + sidebands? cite handbook?} the solution to 
the time-periodic Rice-Mele problem, $[i\dpa_t -H_0(k,t)]\ket{\Y_\n(k,t)}=0$, with the multiple values across the different sideband peaks capturing the indeterminacy of quasienergy up to integer multiples of the drive frequency, $\Omega$. 
The  single particle Floquet states~\cite{Shirley1965,Sambe1973} are given by $\ket{\Y_{\n}(k,t)}=e^{-i\ve_{\n}(k) t}\sum_m e^{-i\W m t}\ket{\f_{\n}^m(k)}$, where $\{\ket{\f_{\n}^m(k)}\}$ are time-independent states. Here, $\ve_{\n}(k)$ is the corresponding quasienergy of the single particle state with crystal momentum $k$, and chirality (or Floquet band) index $\nu = \{\yL, \yR\}$ for the net left- and right-moving bands, respectively. We denote the bandwidth of the Floquet bands by $W_F$ and the gap between them by $\D$.
The net chiralities of the bands are determined by the topological index~\cite{Kitagawa2010,Rudner2020a} $\mathcal{W} = \W\inv \oint dk\, \dpa_k \ve_{R/L}(k)=\pm 1$, where the $+$ ($-$) sign corresponds to the $R$ ($L$) band. The chiralities of the bands are exhibited by the spectral function shown in Fig.~\ref{fig:System}c. As the momentum changes from $k=-\p/a$ to $k=\p/a$, the peaks of the spectral function of the $R$ ($L$) band shift in frequency from $\w$ to $\w\pm \W$.

\section{%Analytical study of the 
Time-evolution towards the quasisteady state\label{sec:Evolution}}

We now study the dynamics of the system in the interacting case. In particular, we focus on the evolution of the two-point Green's functions, providing information about expectation values of one-body operators, such as particle densities and the current (see below). 
 The time-evolution of the interacting system's {two-point} %single-particle
 Green's functions is described by the Kadanoff-Baym equations~\cite{Baym1962}
\begin{subequations}
\begin{eqnarray}
&&[i\dpa_t-H_0(t)] G^{R}(k;t,t')=\delta(t -t')+\Si^R \circ G^R\label{eq:KadanoffBaym_a}\\
&&[i\dpa_t-H_0(t)] G^{<}(k;t,t')=\Si^R \circ G^<+  \Si^< \circ G^A,\label{eq:KadanoffBaym_b}
\end{eqnarray}
\label{eq:KadanoffBaym}%
\end{subequations}
where for brevity we have suppressed the crystal momentum and time indices on the right hand sides of these equations, and the $\circ$ symbol indicates a convolution over time and matrix product in the sublattice indices.
In Eq.~(\ref{eq:KadanoffBaym}), the (flavor-averaged) retarded and  lesser Green's functions are defined as $G^{R}_{ss'}(k;t,t')=-\frac{i}{N}\q(t-t')\sum_{\a=1}^N\overline{\av{\{\hat c_{s,k\a}(t),\hat c\dg_{s',k\a}(t')\}}}_{\ro_0}$, and $G^{<}_{ss'}(k;t,t')=\frac{i}{N}\sum_{\a=1}^N\overline{\av{\hat c\dg_{s',k\a}(t')\hat c_{s,k\a}(t)}}_{\ro_0}$, while $G_{ss'}^A(k;t,t')=G_{s's}^R(k;t',t)\dg$. In these expressions, the expectation values are calculated with respect to the initial state described by the density matrix $\ro_0$, describing an equilibrium state with respect to $\cH(0)$ with temperature $\be_0$ and average density of particles $n_0$. %\addMR{\bf [Is it grand canonical, such that this is the average particle density (not fixed)?]}. 
The bar denotes averaging over the random interaction strength (in the case of the SYK interactions), ``$\bC{,}$'' denotes an anticommutator, and $\q(t)$ is the Heaviside step function. Throughout, we omit the sublattice indices $s,s'$, leaving the $2 \times 2$ matrix structure of the Green's functions implicit. % assuming implicit.} %ly matrix products, unless otherwise specified.}
The retarded and lesser components of self-energy are denoted by $\Si^R(k;t,t')$ and $\Si^<(k;t,t')$, respectively (see
 %include all irreducible diagrams as shown in 
 Fig.~\ref{fig:Interacting}a and Appendix~\ref{sec:Renormalization} for technical details). %\addMR{\bf [MR: time indices in the same order or reversed? Here I'd recommend to show the k and time indices explicitly]}.}

% \cBlue{excluding Hartree-Fock terms, to ensure no spontaneous order is developed, see Fig.~\ref{fig:Interacting}b. The formation of spontaneous order in the system can dramatically change the properties of the system and therefore is beyond the scope of this work. } 

% The many-body evolution of the pump naturally separates into processes on three different time scales. On the shortest time scale, the system follows the renormalized Floquet-Bloch dynamics. These processes include  time-evolution of the renormalized Floquet-Bloch states, without changing their population probabilities \addMR{\bf [MR: What do the previous two sentences mean, especially in the case where the scattering time is short compared with the period? Should we break it down into just two timescales?]}. The intermediate and long time scale processes describe respectively the intraband and interband population dynamics such as thermalization and heating, which is the main focus of this work. To describe the population dynamics we derive a semi-classical kinetic equation. Similar kinetic equations are often used in the studies of weakly interacting dynamical systems [Rammer1986]. Here, we generalize this approach to the strongly interacting case. 

\subsection{Single-particle spectral function \label{sec:SpectralFunction}}%Effective single-particle spectrum}

The renormalized single-particle spectrum of the non-equilibrium system is %short time evolution of the system is 
encoded in the retarded Green's function, whose time-evolution is given by Eq.~\eqref{eq:KadanoffBaym_a}.
In order to facilitate the separation of intraband and interband scattering processes below, we write the full retarded Green's function as a sum of $R$- and $L$-band projected Green's functions: % the intraband and interband scattering processes, we write the retarded function as a sum of band resolved functions, 
$G^R=G_R^R+G_L^R$. 
The band-resolved Green's functions are defined through the Dyson series shown in Fig.~\ref{fig:Interacting}a, corresponding to Dyson's equation, 
\EqS{
&G^R_{\n}(k;t,t')=g^R_{\n}(k;t,t')+ %\\ 
g^R_{\n}\circ \Si^R \circ G^R, %
%&+\int dt_1 dt_2 g^R_{\n}(k;t,t_1) \Si^R(k;t_1,t_2) G^R(k;t_2,t').
\label{eq:InterGreen}
}
where, as in Eqs.~(\ref{eq:KadanoffBaym_a}) and (\ref{eq:KadanoffBaym_b}), we suppress the crystal momentum and time indices on the last term for brevity.
Here, $g^R_{\n}(k;t,t')$ is the non-interacting %single-particle 
retarded Green's function projected to band $\nu$, see Appendix~\ref{sec:BareGreen} for more details.

We define the renormalized band-resolved spectral functions as $G_\n^\D(k; \w,\bar t)=i[G_\n^R(k;\w,\bar t)-G_\n^A(k; \w,\bar t)]$,  where $G_\n^R(k; \w,\bar t)$ is obtained via the Wigner transform of the two-time function, $G_\n^R(k;t,t')$, with $\bar{t} = \frac12(t + t')$, and $G_\n^A(k; \w,\bar t)=[G_\n^R(k;\w,\bar t)]\dg$.
The renormalized band-resolved spectral function %obtains a broadening $\sim \x$ \addMR{\bf [MR: 
broadens due to interactions, with tails extending into the gap that decay approximately as $e^{-\omega/\xi}$. In what follows, we focus on the limit $\x\ll \D$, where the band-resolved spectral functions are well separated in frequency, see Figs.~\ref{fig:Interacting}b,d. This separation of the renormalized Floquet bands in the frequency domain %leads to the suppression of the interband scattering rates (see below), which 
is crucial for obtaining a long-lived quasi steady state in the system (see below).

The renormalization of the single-particle spectral function is caused by the dressing of the non-interacting Floquet bands by virtual electron-hole pair creation and annihilation processes. Our analytical estimate near the quasi steady state (see Appendix~\ref{sec:Renormalization}) suggests that the broadening of $G^\D_L$ in the limit $W_F\ll U\ll \D$ is approximately given by $\x_\yL\eqa -W_F/\ln\bR{\frac{U^2}{\D^2}[f^0_L\bar f^0_L+f^0_R\bar f^0_R]}$, %$\x_L\eqa \frac12 U\sqrt{\half (f^0_L\bar f^0_L+2f^0_R\bar f^0_R)}$, 
where $f^0_L$ and $f^0_R$ are the occupation probabilities of the Floquet bands (see below) and $\bar f^0_\n=1-f^0_\n$.

\subsection{The kinetic equation and population dynamics\label{sec:QuasiSteadyStateOccupation}}

To study the formation and properties of the quasisteady state, %intermediate and long-time processes, 
we define occupation probabilities by parametrizing the lesser Green's function as
\Eq{
G^<(k;t,t') =f_R \circ G^A_R-G^R_R\circ f_R+f_L\circ G^A_L-G^R_L \circ f_L,
\label{eq:ParameterizationGl}
}
where crystal momentum and time indices are suppressed on the right hand side. %Here, all the terms are defined in the time and momentum domain, and the products assume a convolution in time and contraction over the sublattice index. 
By analogy to the case of thermodynamic equilibrium, where the (Fourier-transformed) lesser Green's function and the spectral function are related via the Fermi occupation function, in Eq.~(\ref{eq:ParameterizationGl}) the Hermitian matrices $f_R(k;t,t')$ and $f_L(k;t,t')$ play the roles of distribution functions for the two bands. %the occupation probabilities of the two bands are \addMR{defined in terms of} the Hermitian matrices $f_R(k;t,t')$ and $f_L(k;t,t')$. 
As we will discuss further below, this interpretation is most meaningful when $f_R(k;t,t')$ and $f_L(k;t,t')$ take simple forms in terms of their matrix and time (or frequency) structures. 
We will see that such a simple form naturally emerges in the quasisteady state of the system.

To assess the dynamics of the distribution functions %occupation probabilities
we derive a kinetic equation for $\dpa_{\bar t} f_L(k;\w,\bar t)$, where $ f_L(k;\w,\bar t)$ is obtained via the %resulting from the 
Wigner transform of the two-time function, $f_L(k;t,t')$. %and $\bar{t} = \frac12(t + t')$. %{\bf [MR: okay?]}} %, with respect to the difference in times. 
%\addMR{\bf [MR: what derivative is this referring to? We don't see any derivatives yet...]} and the time-derivative is defined with respect to the average time, $\bar t$. 
A similar approach is often employed in studies of weakly interacting dynamical systems~\cite{Rammer1986,Rammer2007,Kamenev2011}. Here, we generalize this approach to the strongly interacting case, without imposing the ``on-shell'' approximation~\cite{Picano2021}. 
Crucially, such an approximation would not correctly capture interband scattering, which occurs through high-order processes in $U$.
%\addMR{\bf [MR: are we assuming here $W > U,\Omega$?]}} %, and thus can not be approximated by the second-order terms in the interaction.}

The kinetic equation is obtained by combining the Wigner transforms of  Eqs.~\eqref{eq:ParameterizationGl} and \eqref{eq:KadanoffBaym_b} and subtracting terms proportional to $\partial_{\bar{t}} G^R(k; \omega, \bar{t})$, which itself is described by Eq.~\eqref{eq:KadanoffBaym_a} (see Appendix~\ref{sec:KineticEquation} for the full derivation). 
Generically, the kinetic equation for $f_L$ can be written as \cite{Genske2015} $\dpa_{\bar t}f_L(k;\w,\bar t)=\cI_L(k;\w,\bar t)\bC{f_R,f_L}$, where the collision integral $\cI_L$ is a functional of $f_R$ and $f_L$ with a matrix structure in the sublattice indices.
%For the values of $\w$ where $G^\D_L(k;\w,t)$ %\addMR{\bf [MR: $G^R_L$? $G^<_L$?]} 
%has significant weight, i.e., near the lower band (see Fig.~\ref{fig:Interacting}d), the kinetic equation is given by \addMR{\bf [This seems to be a very general form. How does it change further away in $\omega$?]} $\dpa_{\bar t}f_L(k;\w,\bar t)=\cI_L(k;\w,\bar t)\bC{f_R,f_L}$, where the collision integral $\cI_L$ is a functional of $f_R$ and $f_L$ with a matrix structure in the sublattice indices. %The collision integral includes a sum of the transition rates driven by the time-derivatives of the density of states and the self-energy. \addMR{\bf[?]} As a result, the scattering rates are proportional to $\propto \W U$. \addMR{\bf[Really? Rate linear in $U$?]}
Notably, for values of $\w$ where $G^\D_L(k;\w,t)$ has significant weight, i.e., near the lower band (see Fig.~\ref{fig:Interacting}d), the net collision integral is exponentially small if its arguments are independent of $k$ and $\omega$ [see appendix \ref{sec:KineticEquation}]: %for maximally thermalized bands with constant (yet possibly different) occupation probabilities described by $k$ and $\omega$ independent distribution functions 
$f^{\rm (qs)}_L(k; \w,\bar{t})=f_L^0{\bf 1}$, $f^{\rm (qs)}_R(k; \w, \bar{t})=f_R^0{\bf 1}$, %and $\D f\ne 0$, 
where $f_L^0$ and $f_R^0$ are constants and ${\bf 1}$ denotes the identity matrix in sublattice space. In particular, we estimate $\cI_L\bC{ f^{\rm (qs)}_L ,f^{\rm (qs)}_R}  \propto  e^{-\D/\x}\dl f$, where $\dl f =f_R^0-f_L^0$. 
A state described by this form of $f^{\rm (qs)}_\n$, characterized by uniform occupation within each band, exhibits exponentially slow population dynamics and thus describes a long-lived quasisteady state of the system.

\subsection{Universal value of the current}

Using the form of the quasisteady state found above in terms of two-point
%single-particle 
Green's functions, we can now characterize {\it observables} in the quasisteady state.
In particular we focus on the value of the time-averaged current, which was previously conjectured to take a universal value based on a weak-coupling picture and evidence from numerical simulations on modestly-sized systems.
%Once we established the prethermal state in the system, we are in the position to show that the period-averaged current in this state obtains a universal value. 

The instantaneous current averaged along the chain  %\addMR{\bf [MR: this is averaged with respect to position around the ring, right?]} \cBlue{\bf{[Isn't it the same as the local current because we have  translational invariance?]}} 
in a generic translation-invariant state described by $G^<$ reads 
\Eq{
\cJ(t)=-i\int \frac{dk}{2\p} \tr{\dpa_k H_0(k,t)G^<(k;t,t+0^+)},
\label{eq:AverageCurrent}
}
where the momentum integral is performed over the first Brillouin zone. 
Next, we evaluate Eq.~\eqref{eq:AverageCurrent} in the quasisteady state given by $G_{\rm (qs)}^<(k; t, t') = if_R^0 G_R^\D(k; t, t') + if_L^0 G_L^\D(k; t, t') + \mathcal{O}(e^{-\Delta/\xi})$ %\addMR{\bf [MR: okay?]} 
[see Eq.~\eqref{eq:ParameterizationGl} and the following discussion]. 
At equal times, $G_{\rm (qs)}^<$ can be further simplified using $G^\D_\n(k;t,t+0^+)=g^\D_\n(k;t,t+0^+)$, due to the fact that the time-integrals in Eq.~\eqref{eq:InterGreen} vanish for $t=t'$. 
Substituting the resulting quasisteady state form of $G_{\rm (qs)}^<$ into Eq.~\eqref{eq:AverageCurrent}, we obtain $\cJ(t)=f^0_R \cJ_R^0(t)+f^0_L\cJ_L^0(t)$, where $\cJ^0_\n=\int \frac{dk}{2\p} \braoket{\Y_\n}{\dpa_k H_0}{\Y_\n}$ %\addMR{\bf [MR: What is $\Ket{\Psi_\nu}$? Note that $H_0$ is a 2$\times$2 matrix, so if $\Ket{\Psi_\nu}$ is meant to be a many-body state, then this doesn't look quite right. Also, what do we mean by the ``current in the non-interacting Floquet bands?'' Current under what conditions? That they're full?]} 
is the current carried by Floquet band $\nu$ of the system in the absence of interactions, when fully filled (see Appendix~\ref{sec:BareGreen}). As defined in Sec.~\ref{sec:nonInter} above, $\ket{\Y_\n(k,t)}$ denotes the single-particle Floquet state with crystal momentum $k$ in band $\nu$ of system in the absence of interactions.  In the adiabatic limit, the period averaged current $\bar \cJ^0_{\n}=\frac{1}{T}\int_0^T dt \cJ^0_{\n}$  %\addMR{\bf [MR: the current itself isn't ``interacting'' or ``non-interacting.'']} of the non-interacting system 
is quantized~\cite{Thouless1983} as $\bar \cJ^0_{R,L}=\pm\frac{1}{T}$. 
In a system where the upper ($R$) band is initially empty and the lower ($L$) band has fractional filling $\n_0=a n_0$, we therefore expect % assuming a quasi steady state with 
$f_L^0= \n_0$ and $f_R^0= 0$, %we arrive at 
such that the current in the quasisteady state is equal to $\bar\cJ^{(qs)}=\frac{\n_0}{T} + \mathcal{O}(e^{-\Delta/\xi})$, where the correction captures the deviation of the quasi steady state from the maximal entropy state in which the upper band is empty. %\addMR{\bf [MR: + correction?]}.
 
This is a remarkable result: even in a limit where the single particle scattering lifetime may be short compared with the driving period, where the single particle Floquet states and associated spectrum are not well-resolved or defined, the current still attains a universal value associated with the nontrivial topology of the system's single-particle Floquet spectrum in the absence of interactions.
 The universal value of current holds, up to an exponentially small correction, provided that the scattering rate (and associated level broadening, captured by $\xi$) remains small compared with the single particle band gap, $\Delta$.
 %Notice that in the infinite time limit, the system thermalizes reaching $f_L^0=f_R^0=n_0/2$, corresponding to $\bar\cJ=0$.

% The quasi steady state established in the system has important consequences on the period averaged current. While the current in a partially filled pump is expected to be non-universal, it becomes nearly universal in the quasi steady state. To see that, recall that the period-averaged current at $t=nT$ is given by 
% \Eq{
% \bar \cJ(t)=\iint_{t}^{t+T} \frac{dkdt'}{2\p T} \tr{\dpa_k H_0(k,t')G^<(k;t',t'+0^+)}.
% \label{eq:AverageCurrent}
% }
% Using Eq.~\eqref{eq:ParameterizationGl} in the quasi steady state, we find  $G^<(k;t,t+0^+)=f_R^0 (t)G_R^\D(k;t,t+0^+)+f_L^0(t) G_L^\D(k;t,t+0^+)+\cO(\lm)$. This expression can be further simplified using $G^\D_\n(k;t,t+0^+)=g^\D_\n(k;t,t+0^+)$, since the time-integrals in Eq.~\eqref{eq:InterGreen} vanish for $t=t'$. Therefore, the period-averaged current reads $\bar \cJ(t)=f_R^0(t) \bar\cJ_R+f_L^0(t) \bar\cJ_L$, up to order $\cO(\lm)$. Here, $\bar\cJ_\n=\frac{1}{T}\int_0^T dt\int \frac{dk}{2\p} \braoket{\Y_\n}{\dpa_k H_0}{\Y_\n}$ is the single particle current (recall that the single-particle Floquet states $\ket{\Y_\n(k,t)}$ are defined above Eq.~\eqref{eq:BareGreen}), yielding in the adiabatic limit $\bar\cJ_{R/L}=\pm \W$. At early times of the evolution (yet, well after the system reached the quasi steady state), $f^0_L\eqa n_0$ and $f^0_R\eqa 0$, giving rise to $\bar \cJ(t)\eqa n_0\W$.

\begin{figure}
  \centering
  \includegraphics[width=8.6cm]{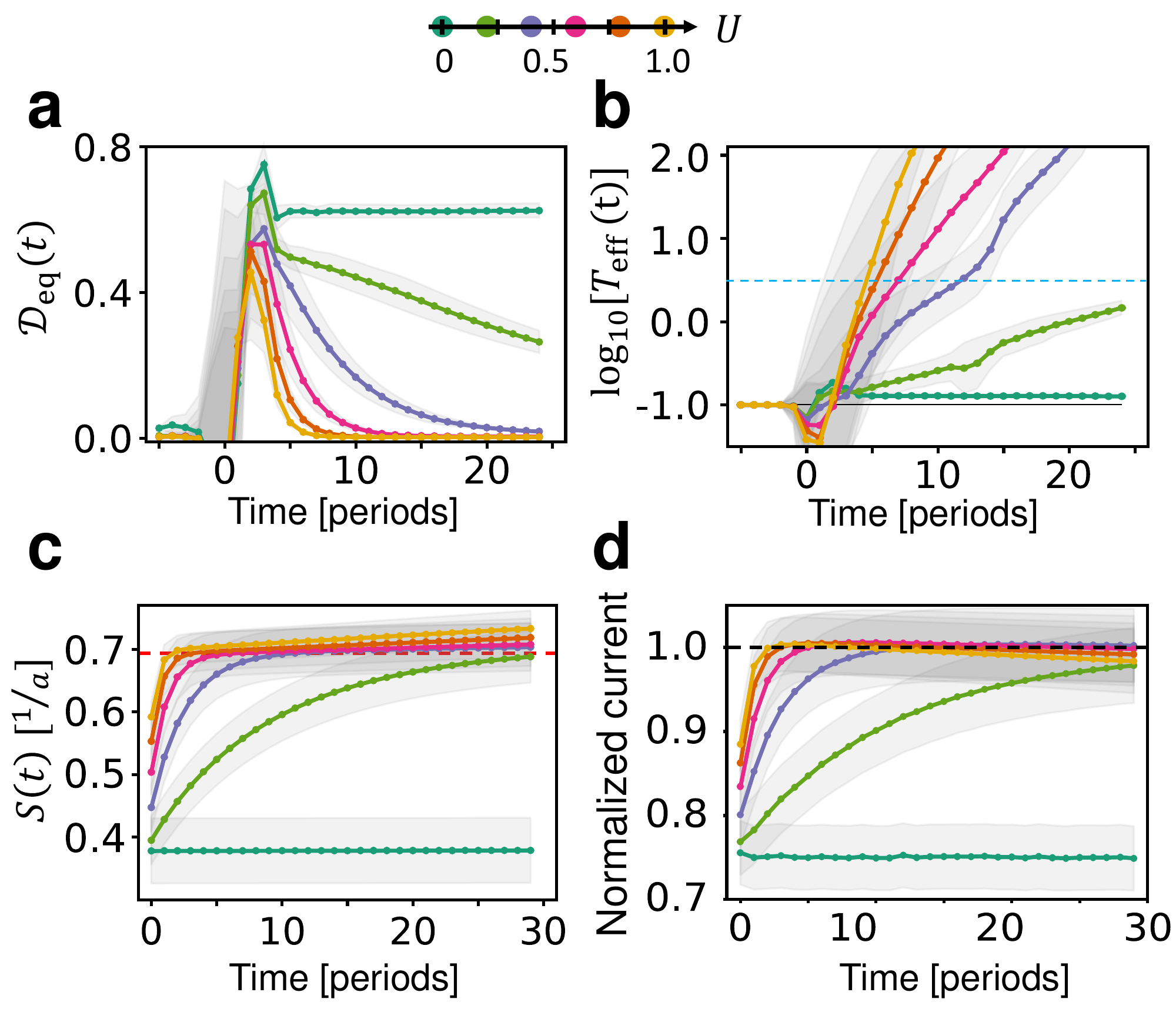}\\
  \caption{ 
\tb{Relaxation to the quasi steady state.} 
\tb{a} ``Distance'' to equilibrium, $\cD_{\rm eq}(t)$, defined in Eq.~\eqref{eq:DistanceEq}, as a function of time. Lines of different colors in this and following panels correspond to the different interaction strengths indicated on the top of the figure, in units of $J_0=1$. Shaded areas %in this and following panels 
indicate error bars due to the extrapolation procedure, see text. At $t=0$, the extrapolation yields non-physical, negative values of $\cD_{\rm eq}(t)$ due to the quench, we thus cut-off these values in the plot.
\tb{b} Effective temperature of the lower band minimizing $\cD_{\rm eq}(t)$ in units of $J_0=1$, as a function of time. Blue dashed line indicates the point where $T_{\rm eff}=W_\yF$.
%\addMR{\bf [What units is $T_{\rm eff}$ in?]}
%Shaded area approximately indicates the region where the system is far from equilibrium, thus effective temperature does not faithfully represent the state of the system. %The effective temperature increases exponentially with the rate $\G_{\rm intra}$. 
\tb{c} Entropy density of the system’s one-body reduced density
matrix %per state 
as a function of time, defined in Eq.~\eqref{eq:Entropy}. A dashed line indicates maximal entropy for a quarter-filled system, when all the particles occupy the lower band. % Effective chemical potential, $\m_{\rm eff}$ in the lower band as a function of time. In the infinite temperature limit $\m_{\rm eff}\to -\infty$. 
%\tb{d} Occupation probability as a function of momentum and time, $f(k,t)$, for $U=0.5$. The system is initially occupied around the center of the Brillouin zone, following equilibrium distribution (see Fig.~\ref{fig:System}b). 
%During the time-evolution, the distribution spreads uniformly across the Brillouin zone, reaching a maximally thermal state
\tb{d} Period averaged current normalized by the total density of particles and driving frequency,  $\bar \cJ(t)(T/\n_0)$ as a function of time.} %The shaded areas represent error bars in the extrapolation procedure, see text.
 \label{fig:Intraband}
\end{figure}
\section{Numerical analysis}

To benchmark our analytical results, we numerically simulated the model given in Eqs.~\eqref{eq:HamiltonianRM} and \eqref{eq:SYKHamiltonian}, with SYK interactions. 
In addition to the terms described above, we also included a weak random quadratic term $\hat\cH_{\rm SYK-2}=\sum_{j,\a\be}K_{\a\be}\hat c_{j\a}\dg\hat c_{j\be}+{\rm h.c.}$,
where $\overline{K_{\a\be}}=0$ and $\overline{K_{\a\be}^2}=K^2/N$, %where we take 
with $K=0.05$.
This additional term is essential for stabilizing the numerics in the weakly interacting regime.

The unique structure of the SYK interactions allows us to simulate considerably larger systems compared to exact diagonalization methods applied to systems with conventional interactions. Here, we simulated the time evolution of a chain of $100$ SYK dots arranged into $L=50$ unit cells. We used the Kadanoff-Baym equations [given in Eq.~(\ref{eq:KadanoffBaym})] to evolve the Keldysh-ordered Green's functions in time~\cite{Stan2009,Eberlein2017,Bhattacharya2019,Kuhlenkamp2020,Maldacena2021}; for further details see Appendix~\ref{sec:Numerics}. The system is initialized in an equilibrium state of $\hat\cH(0)$ with temperature $\be_0\inv=0.1$ and $\m_0=-2.93$, which is set to fix the density of electrons approximately at quarter filling, see Fig.~\ref{fig:System}b.
For the model itself, we select the parameter values: $J_0=1$, $J_1=0.85$, $v_0=2.55$, $\W=0.5$, see Eq.~\eqref{eq:HamiltonianRM} and surrounding text for the definitions of the Hamiltonian and its parameters. We note that all the energies and frequencies are given in units in which $J_0=1$. %{\bf [MR: should we give in the caption? \cBlue{Should we add it to the caption of Fig.~1?} Also, I noticed that it's not easy to find the definitions of $s_0$ and $s_1$ in the text above. Let's think about how to improve this situation.]}}

The time-evolution algorithm is based on the discretization of the time and frequency domains, with small steps $\dl t$ and $\dl \w$, respectively. We performed the evolution for several  values of steps in the range $0.08 \le \delta t \le 0.16$ and $0.015 \le \delta \omega \le 0.04$, and performed a two-dimensional linear extrapolation to $\dl t=\dl \w=0$. In the numerical results we present the extrapolated values with error bars indicating the uncertainty %upper bounds on the error 
 of the extrapolation procedure (defined as the difference between the extrapolated value and the closest numerically-determined point).

\subsection{Formation of the quasisteady state}
We first analyze the %intraband thermalization towards the infinite temperature state of each of the bands.
formation of the quasisteady state, wherein the distribution functions $f_L$ and $f_R$ for the two bands become independent of crystal momentum and frequency, while the total populations of the two bands remain approximately constant.
As a means of characterizing the nonequilibrium state of the system, and to enable the extraction of an effective temperature (when it is appropriate to do so),  we define the ``distance'' of the distribution from a thermal equilibrium state as
\Eq{
\cD_{\rm eq}(\bar{t})=\min_{\be,\m}\int\frac{a\cdot dk }{2\p}\int_{-\infty}^{\w_c}\frac{d\w }{2\p} {\rm Tr}\abs{i\overline G^<+f_{\rm FD}(\be,\m)\overline G^\D},
\label{eq:DistanceEq}
}
where $\w_c$ is the center of the spectrum and the Green's functions %$\overline{G}^<(k;\omega, \bar{t})$ and $\overline{G}^\Delta(k;\omega, \bar{t})$ 
are evaluated in the Wigner transformed representation and averaged over one period: $\overline{G}^<(k;\omega, \bar{t}) \equiv \frac{1}{T}\int_{-T/2}^{T/2} ds\, G^<(k;\omega, \bar{t} + s) $ and similarly for $\overline{G}^\Delta(k;\omega, \bar{t})$. Here, $f_{\rm FD}(\be,\m)=[1+e^{\be(\w-\m)}]\inv$ is the Fermi function. %-Dirac distribution.
For a system in thermal equilibrium, the integrand in Eq.~(\ref{eq:DistanceEq}) vanishes for all $k$ and $\omega$, corresponding to  $\cD_{\rm eq}=0$. 
 
In Figs.~\ref{fig:Intraband}a,b we plot the evolution of the distance to equilibrium $\cD_{\rm eq}(\bar{t})$ and the extracted effective temperature $T_{\rm eff}=\be_{\rm min}\inv$, corresponding to the minimization in Eq.~(\ref{eq:DistanceEq}).
When the drive is switched on at $t=0$, $\cD_{\rm eq}$ rapidly grows, indicating evolution into a far from equilibrium state. 
Following the rapid rise, the system relaxes to the quasisteady state, which is manifested by the decay of $\cD_{\rm eq}$. 
In parallel, the effective temperature grows approximately exponentially with a rate $\G_{\rm intra}$: $T_{\rm eff}\sim \be_0\inv e^{\G_{\rm intra}t}$ (Fig.~\ref{fig:Intraband}b).
The quasisteady state observed in the simulation approximately realizes the conditions discussed in Sec.~\ref{sec:QuasiSteadyStateOccupation}, once the effective temperature exceeds the width of the single-particle Floquet bands, $T_{\rm eff}\gg W_\yF$, indicated by blue dashed line (and for $\mathcal{D}_{\rm eq} \ll 1$). % small {\bf [MR: what are the units of $\mathcal{D}_{\rm eq}$?]})}.}
The curves of different colors in Figs.~\ref{fig:Intraband}a,b correspond to different interaction strengths, $U$; the time to reach the quasisteady state rapidly decreases with interaction strength, $U$.
%We extracted $\G_{\rm intra}$ as a function of $U$, see Fig.~\ref{fig:Interband}b.

To track the system's evolution towards a high entropy density state, % in each of the bands and between the bands, 
we calculated the average von-Neumann entropy density of the system's one-body reduced density matrix %\addMR{\bf [MR/NL: why do we need the $G^>$ term?]} %{\color{red} Is this the same as the entropy calculated from the single-particle density matrix? In that case, shouldn't there be an integral over $\omega$ inside the log?}: % of each single-particle state,  defined as 
\Eq{
S(\bar{t})=-{\rm Tr}\int \frac{ dk}{2\p}\bS{(-i\overline G^<)\log(-i\overline G^<)+(i\overline G^>)\log(i\overline G^>)},
\label{eq:Entropy}
}
where $\bar G^>=-i \bar G^\D+\bar G^<$ and the Green's functions are evaluated at equal times $t = t' = \bar{t}$, see Fig.~\ref{fig:Intraband}c.
%\Eq{
% S(t)=-\int \frac{d\w dk}{(2\p)^2}[f_s\log(f_s)+\bar f_s\log(\bar f_s)]{\rm Tr}[\overline G^\D(k;\w,t)],
% \label{eq:Entropy}
% }
%where $f_s(k;\w,t)=-{\rm Tr}[\overline G^<(k;\w,t)]/{\rm Tr}[\overline G^\D(k;\w,t)]$, and $\bar f_s=1-f_s$.
The value of $S(\bar{t})$ for a maximal entropy density state in a quarter-filled system, subject to the constraint that all the particles occupy the lower band, is given by $S^{\rm max}_L=\log(2)/a$. As can be seen in Fig.~\ref{fig:Intraband}c, the entropy density stabilizes slightly above this value due to a small population excited to the upper band at $t=0$. After stabilizing near $S(t)\eqa S^{\rm max}_L$, the entropy slowly grows further due to interband transitions. In the infinite time limit, we expect $S\to S^{\rm max}\eqa  1.12/a$ corresponding to one quarter filling of the entire system.

% In Fig.~\ref{fig:Intraband}d, we show how the particle distribution $f(k,t)=\int \frac{d\w}{2\p} \tr{\overline G^<(k,\w,t)}$ spreads amongst the momentum states as a function of time. 
% %Fig.~\ref{fig:Intraband}d shows the particle distribution as a function of momentum $f(k,t)=\int \frac{d\w}{2\p} \tr{\overline G^<(k,\w,t)}$.
% At $t=0$, the particles occupy states around the center of the first Brillouin zone, corresponding to the initial Fermi-Dirac distribution, as shown in Fig.~\ref{fig:System}b. 
% After the drive is turned on, the distribution relaxes to a form with uniform probability across the full Brillouin zone.

In Fig.~\ref{fig:Intraband}d, we extracted the period-averaged current normalized by the filling, $\bar \cJ/\n_0$ [see Eq.~\eqref{eq:AverageCurrent}]. As follows from the discussion below Eq.~\eqref{eq:AverageCurrent}, we expect an approximately quantized value in the units of $T\inv$ for the normalized current in the quasisteady state.
Fig.~\ref{fig:Interband}b shows the period-averaged current normalized by $\n_0$ % the numerically extracted value of \addMR{\bf [MR: Current normalized by $n_0$ or $J(t)/n_0$?]} $\bar \cJ(t)/n_0$ 
as a function of the stroboscopic time. %period $t=n T$, with $n=0,1,2,...$.
The gray strips indicate the uncertainty intervals associated with the extrapolation to infinitesimal grid spacing in the simulations, as described above.
In the regime of strong interactions, the average current rapidly increases on a timescale set by $\sim\G_{\rm intra}$.
When the quasisteady state is reached, the current obtains the expected universal value to within the uncertainties of the numerical simulation, as expected. For later times, the current slowly decays with the rate $\G_{\rm inter}$ due to interband heating. In the weakly interacting case, the normalized current remains non-universal for much longer times; for these cases, the quasisteady state was not reached within the time window that we were able to simulate. %\textcolor{blue}{CK: %This behaviour is expected 
Interestingly, slowly-driven Fermi liquids have been shown to persist in non-thermal states for parametrically long timescales~\cite{Kuhlenkamp2020}.
The connection between the slow intraband heating observed here and the mechanism in Ref.~\onlinecite{Kuhlenkamp2020} will be interesting to investigate in future work.
%
%parametrically-long heating and thermalization timescales have been observed for slowly-driven Fermi liquids~\cite{Kuhlenkamp2020}.
%Similar behavior has been observed} for slowly driven Fermi liquids, where heating and thermalization times can be parametrically long~\cite{Kuhlenkamp2020}. Our results therefore establish sizable interactions as a crucial ingredient to stabilize topological pumping.}

\subsection{Interband heating and universal current}
\begin{figure}
  \centering
  \includegraphics[width=8.6cm]{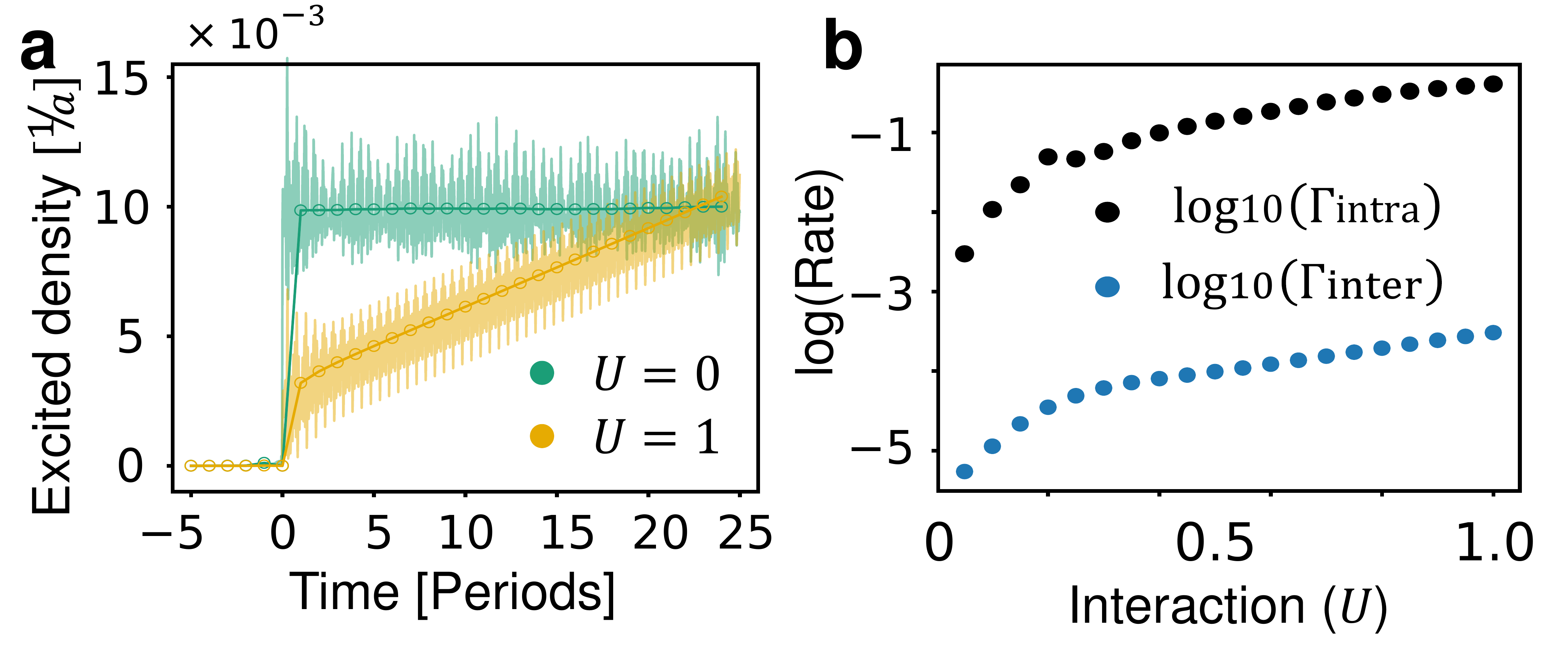}\\
  \caption{%{\color{red} log(Rate) in the y axis label of b?} 
  \tb{Heating and current in the quasi steady state.} \tb{a} Density of electrons excited to the upper band as a function of time, in the interacting and non-interacting cases. Circles represent the period averaged density. 
  In the interacting case, the density of excitations increases with an approximately constant rate $\G_{\rm inter}$, while in the non-interacting case the average charge is constant, following an initial jump at $t=0$.  
\tb{b} Intraband ($\G_{\rm intra}$) and interband ($\G_{\rm inter}$) equilibration rates as a function of the interaction strength. The intraband equilibration rate is extracted from the temperature growth [see Fig.~\ref{fig:Intraband}b]. The interband equilibration rate is extracted from the slope of the period-averaged excitation density, in \tb{a}. 
%  \tb{b} Period averaged current normalized by the total density of particles and driving frequency,  $\bar \cJ(n)/(n_0\W)$ as a function of time, for different interaction strengths indicated on the top of Fig.~\ref{fig:Intraband}. The shaded areas represent error bars in the extrapolation procedure, see text.
\label{fig:Interband}}
\end{figure}

%\addMR{\bf [MR: if the previous section was about the quasisteady state, then it would seem natural to include the current there (rather than in the section about the breakdown of the quasisteady state). What do you think about replacing panel 3d with the current, and then keeping Fig. 4 and this section entirely about interband heating?]}
To investigate the interband scattering processes and measure their rates, in 
Fig.~\ref{fig:Interband}a  we extracted the density of excitations in the (renormalized) $R$ band from our simulations.
We define the excitation density as $n_{\rm ex}(\bar{t})=-i\int_{\omega_c}^\infty \frac{d\w}{2\p}\int\frac{dk}{2\pi}{\rm Tr}\bC{G^<(k;\w,\bar{t})}$. %, where the range of the $\w$-integral is $(\w_c,\infty)$. 
The period-averaged density %\addMR{\bf [MR: should we be using $\overline{G}$ above?] \cBlue{- This sentence talks about the circles in Fig. 4a which are period averaged without the micromotion (i.e., $\overline{G}$ evaluated at the discrete times)}} 
of excitations jumps at $\bar{t}=0$, when the drive is switched on \cite{Privitera2018}, and then gradually increases with an approximately constant rate $\G_{\rm inter}$. 
The interband equilibration rate, $\G_{\rm inter}$, compared to the intraband equilibration rate, $\G_{\rm intra}$, is shown as a function of the interaction strength in Fig.~\ref{fig:Interband}b.

Fig.~\ref{fig:Interacting}c shows the period averaged lesser Green's function as a function of momentum and frequency, after $30$ periods of the drive. The excited population can be seen as a pale strip at the location of the upper band.
Note that the features are heavily broadened due to fast intraband scattering, such that for the selected parameters the Floquet harmonic side bands are nearly completely washed out.

% The discreteness of the frequency domain in the numerical analysis leads to deviations of the calculated normalized charge from the integer value. We thus calculated the current as a function of the discretization step in the frequency domain, $\dl\w$, and extrapolated to $\dl\w=0$. Shaded areas in Fig.~\ref{fig:Interband}c indicate the error of the extrapolation. 

\section{Discussion and outlook}

Periodically driven systems can host Floquet-Bloch bands with unique topological properties that cannot be obtained in equilibrium systems.
In the presence of interactions, it is natural to wonder if rapid scattering on timescales comparable to the driving period might mask any dynamical features expected to arise from the single-particle Floquet states.
In this work we showed that this need not be the case: even in a state with high entropy density and rapid scattering, universal transport associated with the topological properties of the system's Floquet Bloch bands persists.
We demonstrated this phenomenon in the context of a topological pump with non-integer filling, which exhibits a long-lived quasisteady state with maximal entropy density (subject to the constraint of fixed particle number in each band).
We derived conditions under which the quasisteady state hosts quantized transport (in units of the particle density), up to an exponentially small correction in the ratio of the system's band gap to its renormalized band width.

To support these arguments, we studied this phenomenon numerically in an SYK-type chain.
This setup enabled us to examine the dynamics in a regime of strong scattering, in system sizes much larger than could be accessed by exact evolution.
This advantage is gained from the fact that the SYK system can be solved by time evolution of the Kadanoff-Baym equations.
Our numerical results allowed us to study the dynamics leading to the formation of the quasisteady state. Importantly, we showed that quantized transport persists even when quasiparticles are short-lived due to fast intraband scattering and the Floquet sidebands are hence not well resolved.

\section{Acknowledgments}
We would like to thank Ervand Kandelaki and Michael Knap for illuminating discussions, %and assistance with setting up the numerical simulations, 
and David Cohen and Yan Katz for technical support. N. L. acknowledges support from the European Research Council (ERC) under the European Union Horizon 2020 Research and Innovation Programme (Grant Agreement No. 639172), and from the Israeli Center of Research Excellence (I-CORE) “Circle of Light”. M. R. gratefully acknowledges the support of the European Research Council (ERC) under the European Union Horizon 2020 Research and Innovation Programme (Grant Agreement No. 678862) and the Villum Foundation. M. R. and E. B. acknowledge support from CRC 183 of the Deutsche Forschungsgemeinschaft. G.R. acknowledges support from the U.S. Department of Energy, Office of Science, Basic Energy Sciences under Award desc0019166 and the Simons Foundation.

%\bibliography{Bibliography}

\appendix

\section{The band-resolved bare Green's function \label{sec:BareGreen}}
Here, we present the derivation of the retarded band-resolved bare Green's function, see Fig.~\ref{fig:System}c,d. In the non-interacting case the flavors (denoted by $\a$ in Eq.~\eqref{eq:SYKHamiltonian}) are independent of each other. We thus focus on $\a=1$ and omit the flavor index. The bare retarded Green's function is defined as 
\Eq{
g^R_{ss'}(k;t,t')=-i\av{\{\hat c_{s,k}(t),\hat c_{s',k}\dg(t')\}}\q(t-t'),
\label{eqA:RetardedGreenDefinition}
}
where $s,s'=\{A,B\}$ are the sublattice indices, and $\hat c_{s,k}(t)=\hat U\dg (t)\hat c_{s,k}(0) \hat U(t)$. The unitary evolution operator is given by $\hat U(t)=\cT e^{-i\int_{-\infty}^t\hat \cH(t')dt'}$, where $\hat \cH(t)$ is given in Eq.~\eqref{eq:HamiltonianRM}. For $t$ and $t'$  well after the quench, the time-dependent Hamiltonian can be diagonalized, by the Floquet eigenstates $\ket{\Y_\n(k,t)}$, for $\n=R,L$. In this eigenbasis the fermionic operators read
\Eq{
\hat c_{s,k}(t)=\braket{s}{\Y_R(k,t)}\hat d_{R,k}+\braket{s}{\Y_L(k,t)}\hat d_{L,k},
\label{eqA:FloquetOperator}
}
where $\hat d_{\n,k}$ annihilate the Floquet state $\ket{\Y_\n(k,t)}$ and $\braket{s}{\Y_\n(k,t)}$ is the amplitude of the Floquet state projected onto a sublattice $s$.
Substituting Eq.~\eqref{eqA:FloquetOperator} in Eq.~\eqref{eqA:RetardedGreenDefinition}, and evaluating $\av{\{\hat d\dg_{\n,k},\hat d_{\n',k}\}}=
\dl_{\n\n'}$, we arrive at 
\EqS{
&g^R_{ss'}(k;t,t')=-i\q(t-t')\times \\
&\bS{\braket{s}{\Y_R(k,t)}\braket{\Y_R(k,t')}{s'} +\braket{s}{\Y_L(k,t)}\braket{\Y_L(k,t')}{s'}}.
\label{eqA:RetardedGreen}
}
Following Eq.~\eqref{eqA:RetardedGreen}, we define the right/left chirality Green's functions as
\EqS{
&g^R_{\n}(k;t,t')=-i\q(t-t')\ket{\Y_\n(k,t)}\bra{\Y_\n(k,t')}.
\label{eqA:RetardedChirality}
}
Note, that $g^R_{\n}(k;t,t')$ is essentially a projector to one of the Floquet bands and therefore has a matrix structure in the sublattice indices. 
The original Green's function [defined in Eq.~\eqref{eqA:RetardedGreenDefinition}]  is given by the sum of the band-resolved Green's functions,
$g^R_{ss'}(k;t,t')=\braoket{s}{g^R_R(k;t,t')+g^R_L(k;t,t')}{s'}$.

\subsection{Wigner representation of the retarded Green's function}

Next, we derive the Winger-transformed representation of the band-resolved Green's function, given in Eq.~\eqref{eqA:RetardedChirality}.
The Wigner transform of $g(k;t,t')$ is defined as 
\Eq{
g(k;\w,\bar t)=\int e^{i\w\dl t}g(k;\bar t+\dl t/2,\bar t-\dl t/2)d\dl t .
\label{eqA:WignerTransform}
}
To evaluate Eq.~\eqref{eqA:WignerTransform}, we substitute the harmonic expansion of the Floquet states, $\ket{\Y_{\n}(k,t)}=e^{-i\ve_{\n}(k) t}\sum_m e^{-i\W m t}\ket{\f_{\n}^m(k)}$ (see Sec.~\ref{sec:Evolution}) in Eq.~\eqref{eqA:RetardedChirality}. We then perform the $\dl t$ integral yielding \cite{Kitagawa2011}
\Eq{
g^R_{\n}(k;\w,\bar t)=\sum_{m,n}\frac{\ketbra{\f_{\n}^{m+n}(k)}{\f^{m-n}_{\n}(k)}}{\w-\ve_{\n}(k)+m\W+i0^+}e^{-2in\W \bar t}.
\label{eqA:BareGreen}
}
The period averaged Green's function [cf. Fig.~\ref{fig:System}c] can be extracted from the $n=0$ terms in Eq.~\eqref{eqA:BareGreen}, yielding
\Eq{
\bar g^R_{\n}(k;\w)=\sum_{m}\frac{\ketbra{\f_{\n}^{m}(k)}{\f^{m}_{\n}(k)}}{\w-\ve_{\n}(k)+m\W+i0^+}.
\label{eqA:BareGreenAverage}
}

\section{Definition of the Keldysh-Floquet Green's functions in the gauge-invariant form}

Due to the non-equilibrium nature of the Floquet-Keldysh Green's functions, the energy in the collision processes is only conserved modulo $\W$. This property complicates the calculations using the Keldysh formalism, as multiple photon absorption/emission processes have to be take in account in each collision. 
Here, we present a gauge invariant definition of the Green's functions in which the $\w$ index is conserved in the collision processes, and show how convolution and product of the the two-time Green's functions are defined with this gauge choice. Such a definition, allows us to operate the Keldysh-Floquet Green's functions as equilibrium Keldysh Green's functions with additional matrix structure in the Floquet harmonics. 

Given the Wigner-transformed Green's function $G(\w,\bar t)$ [see Eq.~\eqref{eqA:WignerTransform} for definition], we define the Green's function in the harmonic basis (with indices $m,n\in \dZ$) as 
\Eq{
G_{mn}(\w)=\frac{1}{T}\int_0^T d\bar t G(\w+\frac{m+n}{2}\W,\bar t)e^{i\W(m-n)\bar t}.
\label{eqA:GreenHarmonic}
}
This definition is invariant under the transformation $G_{m,n}(\w+l\W)=G_{m+l,n+l}(\w)$ for any integer $l$. A convolution $C(t,t')=\int ds A(t,s)B(s,t)$, reads
% \Eq{
% C_{mn}(\w)=
% }

% \subsection{Product and convolution of two-point functions}
% Here, we give formulas for convolution and product of Wigner-transformed two-point functions of the form $A(\w,t')$ [see e.g., Eq.~\eqref{eqA:BareGreen}]. For convenience, we define the following matrix in the harmonic indices 
% \Eq{
% A_{mn}(\w)=\frac{1}{T}\int_0^T dt' A(\w+n\W,t')e^{i(m-n)\W t'}.
% }
% Using this definition a time convolution of two functions, $C(t,t')=\int  A(t,\bar t)B(\bar t,t')d\bar t$ is given by
\Eq{
C_{mn}(\w)=\sum_l A_{ml}(\w)B_{ln}(\w). 
\label{eqA:Convolution}
}
Similarly, a product of two same-time functions, $P(t,t')=A(t,t')B(t,t')$ is given by
\Eq{
P_{mn}(\w)=\sum_l \int\frac{d\w'}{2\p}A_{m-l,n}(\w-\w')B_{l,0}(\w').
\label{eqA:Product}
}

\section{Evaluation of the self-energy and renormalization of the spectral function\label{sec:Renormalization}}

Here, we estimate the broadening $\x$ of the renormalized  bandwidth of the Floquet bands, discussed in Sec.~\ref{sec:SpectralFunction}. In all the expressions in this section, we assume the Green's functions in the frequency domain are defined as matrices in the harmonic basis [as defined in Eq.~\eqref{eqA:GreenHarmonic}], and implicitly contract the harmonic indices following the rules given in Eqs.~\eqref{eqA:Convolution} and \eqref{eqA:Product}. The renormalization of the spectral width can be understood from the definition of the renormalized Green's function in terms of the bare one, 
\Eq{
[G^R(\w)]\inv=[g^R(\w)]\inv-\Si^R(\w),
\label{eqA:RenormRetarded}
}
following from Eq.~\eqref{eq:KadanoffBaym_a}.
Focusing on the values of $\w$ at the gap of the bare function, i.e., where $g^\D(\w)=0$, the renormalized spectral function reads 
\Eq{
G^\D(\w)\eqa -[g^R(\w)]^2\Si^\D(\w).
\label{eqA:SpectrumInGap}
}
Therefore, to estimate the broadening of the spectral function, we need to estimate $\Si^\D(\w)$.
In what follows, we estimate the lesser and greater components of the self-energy $\Si^\lessgtr(\w)$, constituting the spectral component, $\Si^\D=i\Si^>-i\Si^<$.

\begin{figure}
  \centering
  \includegraphics[width=8.6cm]{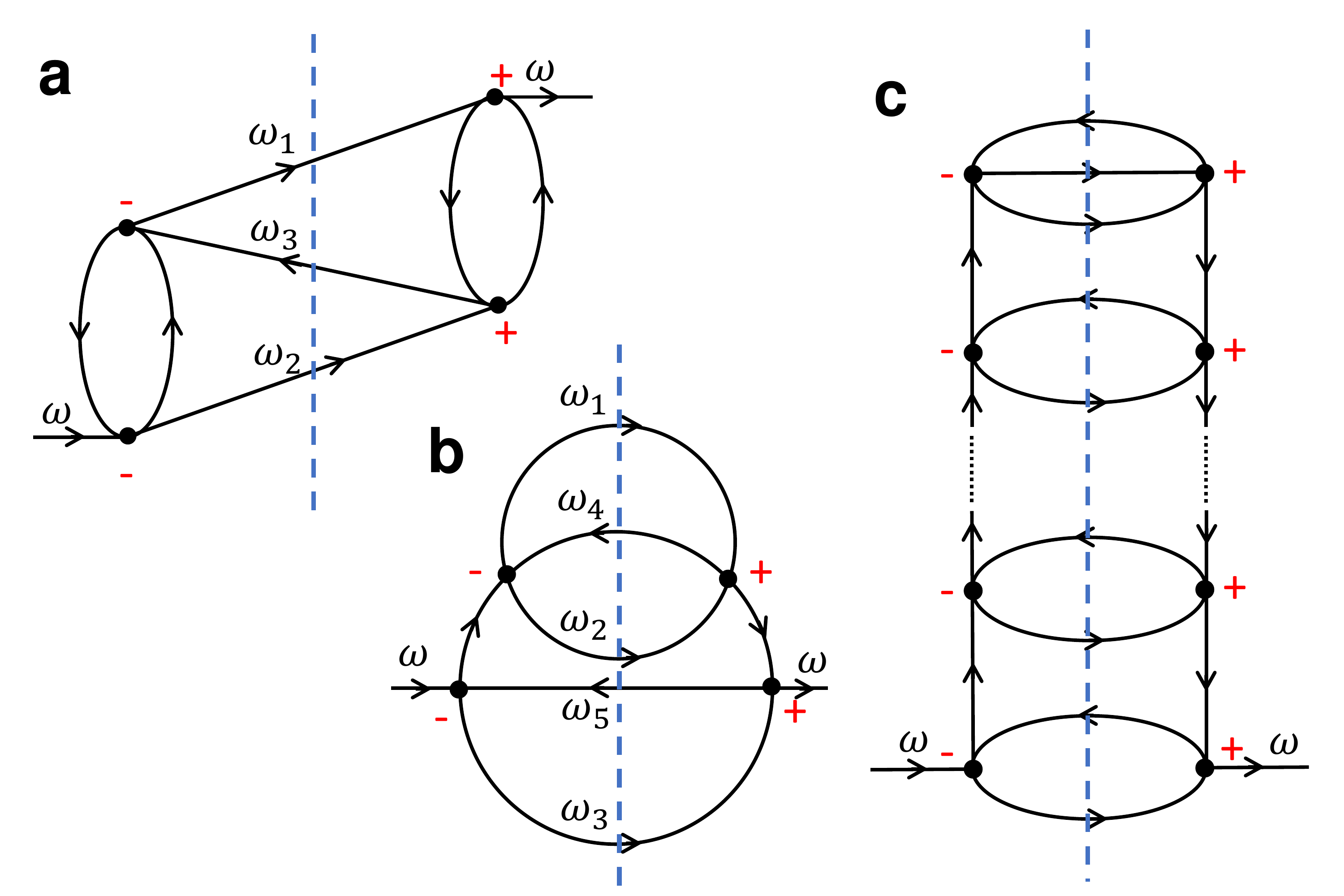}\\
  \caption{Examples of diagrams contributing to the self energy, $\Si^>(\w)$, arranged such that the vertices on the positive Keldysh branch are at the left side and the vertices on the negative Kelysh branch are at the right side. Notice that the left diagram would not contribute in the case of the SYK interactions, since it is subleading in the number of the SYK flavours. \label{fig:SelfEnergy}}
\end{figure}

To estimate the self-energy, we need to sum over all irreducible diagrams allowed by the interaction term (first four terms in the expansion are demonstrated  Fig.~\ref{fig:Interacting}a). We begin by estimating the greater component of the self-energy, $\Si^>(\w)$.
Consider a generic irreducible diagram in this sum, corresponding to the order $U^p$ in the interaction strength, see Fig.~\ref{fig:SelfEnergy}. Such a diagram contains $p$ interaction vertices evaluated at the times $t_1$, $t_2$,.., $t_p$ and positions on the Keldysh contour $i_1,..,i_p$, where $i_p=\pm$ corresponding to positive ($+$) and negative ($-$) branches on the Keldysh contour. The vertices are connected by the non-interacting propagators $g_{i,i'}(t,t')$, with the convention $g_{+-}=g^<$, $g_{-+}=g^>$, $g_{++}=g^T$ and $g_{--}=g^{\tilde T}$.
For a specific combination of $i_p$, it is useful to arrange the diagram such that all the vertices on the positive branch are on the left side and all the vertices on the negative branch are on the right side, see Fig.~\ref{fig:SelfEnergy}. Notice that by the definition of $\Si^>$ the incoming vertex belongs to the negative Keldysh branch and the outgoing vertex belongs to the positive branch.

Next, we transform the expression for the self energy following the transformation given in Eq.~\eqref{eqA:GreenHarmonic}. For convenience, we enumerate the frequencies of the $l$ propagators going from the left to the right side of the diagram by $\w_1$, $\w_2$,..., $\w_l$, and $l-1$ propagators going from the right side to the left side by $\w_{l+1}$, $\w_{l+2}$,..., $\w_{2l-1}$. The maximal value of $l$ is limited by $p$. 
We define the sum of the frequencies of the propagators crossing the center of the diagram with opposite signs by $\D\w=\w_1+...+\w_l-\w_{l+1}-...-\w_{2l-1}$. From the conservation of frequency, $\D\w=\w$, where $\w$ is the frequency associated with self-energy.

By construction, the propagators directed from the left to right correspond  to the greater component of the Green's function, $g^>(\w_1)$,..., $g^>(\w_l)$ while the ones directed oppositely correspond to the lesser components of the Green's function, $g^<(\w_{l+1})$,..., $g^<(\w_{2l-1})$. For simplicity, we suppressed the momentum dependence and the Floquet harmonic indices, as they are not important for the qualitative result. 
Therefore, the greater component of the self-energy of a single diagram to the order $U^p$ is given by 
\EqS{
\Si_p^>(\w)=&U^p\int \frac{d\w_1...d\w_{2l-1}}{(2\p)^{2l-1}}\dl(\w-\D\w)\times\\
&\times g^>(\w_1)\dotsb g^<(\w_{2l-1}) \cF_p(\w_1,...,\w_{2l-1}).
\label{eqA:SelfEnergyEst}
}
The function $\cF_p$ includes the contribution from all the propagators that do not cross the center of the diagram, which include the time-ordered and anti time-ordered components. It is analytical and non vanishing, therefore can not change the support of the convolution in the $\w$ domain, yet it may change the weight of the function.

%While explicit expression of $\cF$ depends on the topology of each particular diagram, it is not crucial for the qualitative estimation of the spectral width of the self energy, provided it is analytical and non-vanishing except of discrete points. 

Our goal is to estimate $\Si^>(\w)$ for $\w$ away from the support of the non-interacting density of states. The dominant contribution to the self-energy would arise from ladder-shaped diagram, as is shown in Fig.~\ref{fig:SelfEnergy}c. Such a diagram has maximal number of lines crossing the center of the diagram, with minimal constraints on the intermediate frequencies.

The unique topology of this diagram allows to write it as a recursive relation \EqS{
\Si^>_{p+2}(\w)=&U^2\int\frac{d\w_1d\w_2 d\w_3}{(2\p)^2}\dl(\w_3-\w+\w_1-\w_2)\times \\
&\times g^>(\w_1)g^<(\w_2)g^T(\w_3)g^{\tilde T}(\w_3)\Si^>_{p}(\w_3).
\label{eqS:SelfEnergyRecursion}
}
with the initial condition
\EqS{
\Si^>_{2}(\w)=&U^2\int\frac{d\w_1d\w_2}{(2\p)^2}g^>(\w_1)g^<(\w_2)g^>(\w-\w_1+\w_2).
\label{eqS:SelfEnergyTermination}
}
For simplicity, we assume a nearly quasi steady state in which the lesser and greater functions can be approximated by $g^>(\w)\eqa i\bar f^0_R  g^\D_R(\w)+i\bar f^0_L g^\D_L(\w)$ and $g^<(\w)\eqa -if^0_R  g_R^\D(\w)-if^0_L g_L^\D(\w)$. The dominant contribution arises from diagrams where each pair of propagators in Eqs.~\eqref{eqS:SelfEnergyRecursion} and \eqref{eqS:SelfEnergyTermination} corresponds to the same band, i.e., $g^>(\w_1)g^<(\w_2)\to (f_L\bar f_L+f_R\bar f_R)g_0^\D(\w_1)g_0^\D(\w_2)$, 
where $g_0^\D(\w)$, is the density of states of one of the  bands shifted to the center of the energy. 
We also approximate $g^T(\w)g^{\tilde T}(\w)\eqa 1/\w^2$. 

The solution to Eq.~\eqref{eqS:SelfEnergyTermination}, can be written as $\Si_2^>(\w)\eqa iU^2(f_L\bar f_L+f_R\bar f_R)[\bar f_R \tilde g^\D_R(3W_F,\w)+\bar f_L \tilde g^\D_L(3W_F,\w)]$, where $\tilde g^\D_\n(d,\w)$ is a function of width $d$ centered around the $\n$-th band. Applying Eq.~\eqref{eqS:SelfEnergyRecursion}, we obtain $\Si^>_4(\w)\eqa i\frac{U^4}{((3/2)W_F)^2}(f_L\bar f_L+f_R\bar f_R)^2[\bar f_R \tilde g^\D_R(5W_F,\w)+\bar f_L \tilde g^\D_L(5W_F,\w)]$.
Similarly, after $p$ iterations, we arrive at
$\Si^>_{2p}(\w)\eqa \frac{iU^{2p}}{(p!)^2 W_F^{2p-2}}(f_L\bar f_L+f_R\bar f_R)^p[\bar f_R \tilde g^\D_R((2p+1) W_F,\w)+\bar f_L \tilde g^\D_L((2p+1) W_F,\w)]$. 
We separate the self energy to $\Si^>_{2p}(\w)=\Si^>_{L,2p}(\w)+\Si^>_{R,2p}(\w)$ where
$\Si^>_{\n,2p}(\w)\eqa \frac{iU^{2p}}{(p!)^2 W_F^{2p-2}}(f_L\bar f_L+f_R\bar f_R)^p \bar f_\n \tilde g^\D_\n((2p+1) W_F,\w)$.

For a distance $|\w|$ from the left band, the leading order in $p$ is proportional to $p\sim |\w/W_F|$. Therefore, the self energy to this order reads 
$\Si_L^>(\w)\eqa \frac{i U \bar f_L}{(\lfloor|\w/W_F|\rfloor!)^2} \bS{\frac{U^2}{W_F^2}(f_L\bar f_L+f_R\bar f_R)}^{|\w|/W_F}$.
Using the Stirling's approximation $p!\eqa p^p$, we obtain $\Si_L^>(\w)\eqa i U \bar f_L \bS{\frac{U^2}{\w^2}(f_L\bar f_L+f_R\bar f_R)}^{|\w|/W_F}$. For $\w$ of the order of $\D$, we arrive at $\Si_L^>(\w)\propto iU\bar f_L e^{-|\w|/\x}$, where 
\Eq{
\x\eqa -W_F/\ln\bR{\frac{U^2}{\D^2}[f_L\bar f_L+f_R\bar f_R]}.
}
Similarly, the lesser component of the self energy reads $\Si_L^<(\w)\propto -iU f_L e^{-|\w|/\x}$. Therefore, $\Si^\D=i\Si^>-i\Si^<\propto Ue^{-|\w|/\x}$. 
Using  Eq.~\eqref{eqA:SpectrumInGap},  we obtain
\Eq{
G_L^\D(\w)\propto e^{-|\w|/\x}.
\label{eqA:SpectrumScaling}
}
A similar calculation near the upper band leads to the same energy scale for the broadening.

\section{Kinetic equation\label{sec:KineticEquation}}
In this section, we derive the kinetic equation for the occupation probabilities, defined in Eq.~\eqref{eq:ParameterizationGl}, and demonstrate that this equation is solved by constant occupations of the bands $f_L=f_L^0$ and $f_R=f_R^0$, up to terms proportional to $\cO(\dl f e^{-\D/\x})$, where $\dl f=f_L^0-f_R^0$. This means that the fixed point of the kinetic equation to the order $e^{-\D/\x}$, corresponds to an infinite temperature distribution in each of the bands. If these terms are included,  the fixed point is a global infinite temperature state, in which $\dl f=0$.

To derive the kinetic equation , we substitute Eq.~\eqref{eq:ParameterizationGl}, in Eq.~\eqref{eq:KadanoffBaym_b}. Before, performing the substitution, we rewrite Eq.~\eqref{eq:KadanoffBaym_b} in the frequency-momentum domain by performing the Wigner transformation of the time-frequency domain, yielding 
\EqS{
&[\w-H_0(k,t)\fatsemi G^<(k;\w,t)]=\\
&=\Si^R\circ G^<-G^<\circ \Si^A-G^R\circ \Si^<+\Si^<\circ G^A.
\label{eqA:EvolutionGl}
}
Here, ``$\circ$'' denotes the Moyal and matrix product and $[A\fatsemi B]=A\circ B-B\circ A$.
To the first order in the derivatives and commutators the Moyal commutator reads, $[A\fatsemi B]=[A,B]+i \dpa_t A \dpa_\w B-i\dpa_w A \dpa_t B$. 
Our goal is to derive an equation for $\dot f_L$ and $\dot f_R$ without imposing the ``on-shell'' approximation, which otherwise would not include the interband transitions occurring off-shell. Eq.~\eqref{eqA:EvolutionGl} describes the time-evolution of $G^<$, which includes time-evolutions of both the spectral function and the occupations [see Eq.~\eqref{eq:ParameterizationGl}]. On the other hand, the evolution of $G^\D$ alone can be derived from Eq.~\eqref{eq:KadanoffBaym_a}, and reads
\EqS{
&[\w-H_0(k,t)\fatsemi G^\D(k;\w,t)]=\\
&=\Si^R\circ G^\D-G^\D\circ \Si^A-G^R\circ \Si^\D+\Si^\D\circ G^A.
\label{eqA:EvolutionGd}
}

To separate the kinetic equation for $f_L$ from the kinetic equation for $G^\D$, we define $\cI_L(k;\w,t)\eqv [\w-H_0\fatsemi G^<]+f_L\circ [\w-H_0 \fatsemi G^\D]$. 
Evaluating the l.h.s. of Eqs.~\eqref{eqA:EvolutionGl} and \eqref{eqA:EvolutionGd}, we obtain
\Eq{
\cI_L=[\w-H_0\fatsemi G^<+f_L\circ G^\D]-[\w-H_0\fatsemi f_L]\circ G^\D.
\label{eqA:CollisonLHS}
}
To simplify, we rewrite Eq.~\eqref{eq:ParameterizationGl} as $G^<=-f_L\circ G_L^\D-f_R\circ G_R^\D+\dl G^<$, where $\dl G^<=[f_L\fatsemi G_L^R]+[f_R\fatsemi G^R_R]$. Therefore, $G^<+f_L\circ G^\D=\dl f\circ G_R^\D+\dl G^<$.
To simplify even further, we assume a close to steady state, such that we can keep only the first order terms in the derivatives and commutators of $f_\n$. With this assumption $\dl G^<$ is already given to the leading order and its Moyal commutator will be of higher order and thus can be neglected. In addition, focusing on $\w\eqa -\D/2$, the spectral function scales as $G^\D_R\eqa e^{-\D/\x}$ [see Eq.~\eqref{eqA:SpectrumScaling}]. Therefore, this term can be neglected compared to $f_L\circ G_L^\D$. With these approximations and explicitly computing the Moyal commutator of $f_L$ in Eq.~\eqref{eqA:CollisonLHS} to the leading order, we obtain, 
\Eq{
\cI_L=(i\dot f_L+i\dot H_0 \dpa_\w f_L-[H_0,f_L]) G_L^\D.
\label{eqA:KineticLHS}
}
Eq.~\eqref{eqA:KineticLHS} is essentially the l.h.s. of the Boltzmann-like equation (up to the band-renormalization terms discussed below). 

Similarly, we evaluate $\cI_L$ using the r.h.s. of Eqs.~\eqref{eqA:EvolutionGl} and \eqref{eqA:EvolutionGd}, yielding the collision term (and band-renormalization terms). 
An explicit calculation yields,
\EqS{
&\cI_L=G^<\circ \Si^\D-G^\D\circ \Si^<+\\
&+[\Si^R\fatsemi G^<+f_L\circ G^\D]-
[\Si^R\fatsemi f_L]\circ G^\D+\\
&+[\Si^<+f_L\circ\Si^\D\fatsemi G^A]-
[f_L\fatsemi G^A]\circ \Si^\D.
\label{eqA:KineticRHS}
}
Evaluation of Eq.~\eqref{eqA:KineticRHS} in a generic state is complex and is performed in the numerical part. As a first order check, we will verify that an infinite temperature state for each of the bands $f_\n=f_\n^0$, with $f^0_L\ne f^0_R$ almost nullifies the collision integral and estimate the timescale for the full thermalization of the system to an infinite temperature state (in which $f^0_R=f^0_L$). Under the assumption of constant occupations, Eq.~\eqref{eqA:KineticRHS} simplifies to
\EqS{
\cI_L=&\dl f  (G_R^\D \circ\Si_L^\D-G_L^\D\circ \Si^\D_R)+\\
&+[\Si^R\fatsemi \dl f  G^\D_R]+[\dl f  \Si^\D_R\fatsemi G^A].
\label{eqA:KineticRHS2}
}
Here, we denote by $\Si_{L}(\w)$ the self energy near $\w=-\D/2$, and similarly $\Si_R(\w)$, denotes the self-energy near $\w=\D/2$. We also used  $\Si_L^<\eqa -if^0_L\Si^\D_L$ and $\Si_R^<\eqa -if^0_R\Si^\D_R$.
Importantly, all the terms in $\cI_L$ are proportional to $\D f$ and to either $G_R^\D$ or $\Si_R^\D$, that are exponentially small near $\w=-\D/2$. This exponentially small value of $\cI_L$ is proportional to the rate of thermalization to the infinite temperature state, in which $\dl f=0$. In this state Eq.~\eqref{eqA:KineticRHS2} becomes identically zero, $\cI_L=0$.

\begin{figure}
  \centering
  \includegraphics[width=8.6cm]{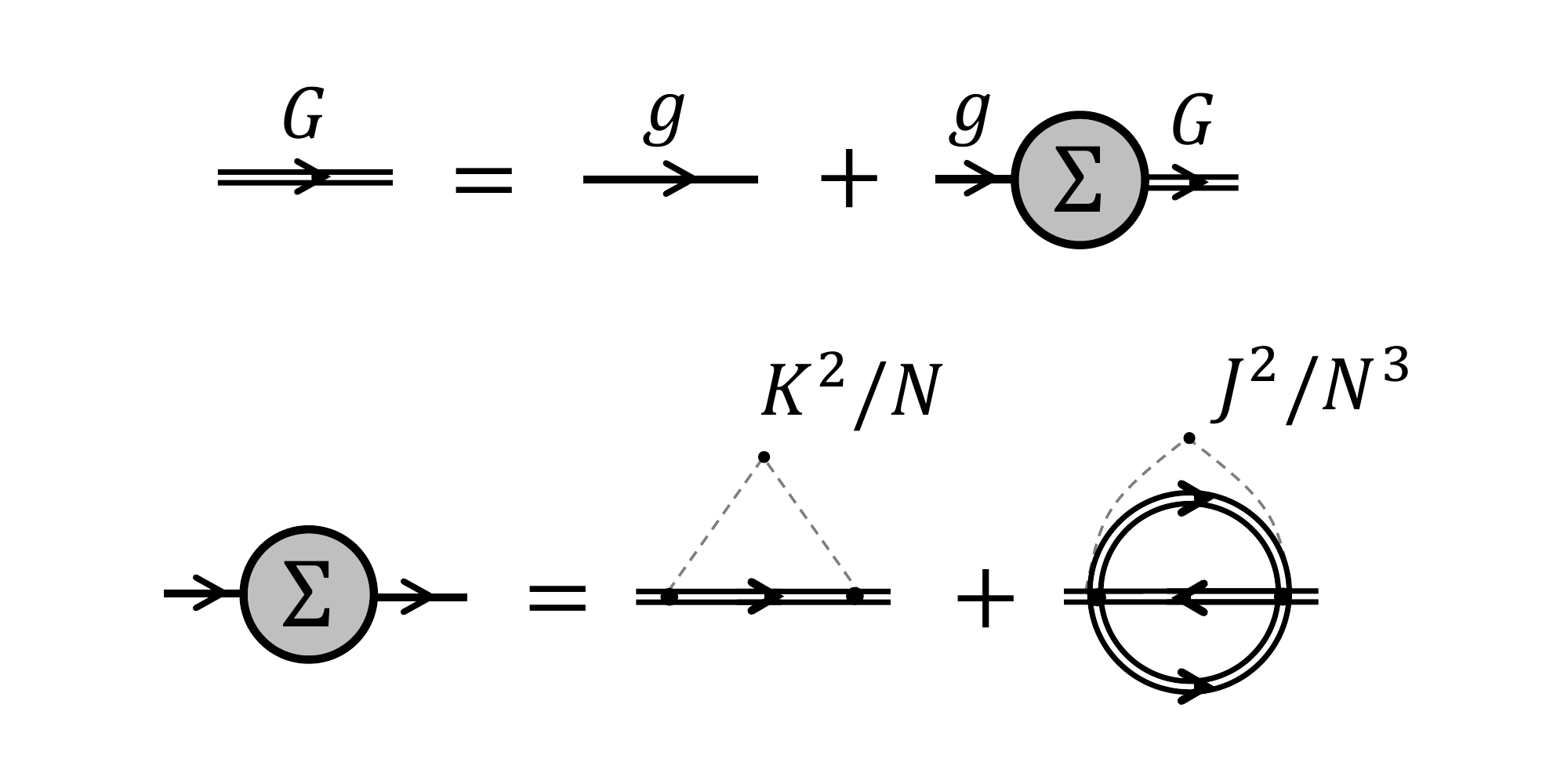}\\
  \caption{Self consistent relations for a Thouless pump with SYK interactions. Single lines represent the non-interacting Green's function $g_{ss'}(k;t,t')$. Double lines represent the renormalized Green's functions $G_{ss'}(k;t,t')$. \label{fig:SYKRelations}}
\end{figure}

\section{Details of the numerical simulation\label{sec:Numerics}}

Here, we present the details of the numerical simulation of the time-evolution of the Kadanoff-Baym equations [see Eq.~\eqref{eq:KadanoffBaym}]. The time-evolution is performed with respect to the SYK Hamiltonian, given in Eq.~\eqref{eq:SYKHamiltonian}, with $N\to \infty$. To stabilize the numerics in the
weakly-interacting limit, we added  a weak random quadratic term 
\Eq{
\hat\cH_{\rm SYK-2}=\sum_{j,\a\be}K_{\a\be}\hat c_{j\a}\dg\hat c_{j\be}+{\rm h.c.},
\label{eqA:HamiltonianSYK2}
}
where $\overline{K_{\a\be}}=0$ and $\overline{K_{\a\be}^2}=K^2/N$.

% This problem admits an exact numerical solution in the limit $N\to \infty$.
% To stabilize the numerics in the weakly-interacting limit, we add a small local disorder term. Such a term induces an inter-chain hopping in each site and is given by
% \Eq{
% \hat\cH_{\rm dis}=\sum_{j,\a\be}K_{\a\be}\hat c_{j\a}\dg\hat c_{j\be}+{\rm h.c.},
% \label{eqA:DisorderHamiltonian}
% }
% where $\overline{K_{\a\be}}=0$, $\overline{K_{\a\be}^2}=K^2/N$, and $K$ is a small constant.  

% We calulate the time-evolution can be analyzed using the disorder-averaged lesser and greater functions \cite{Eberlein2017,Bhattacharya2019,Maldacena2019,Kuhlenkamp2020}, defined as 
% \begin{subequations}
% \begin{eqnarray}
% &&G^{<}_{ij}(t,t')=\frac{i}{N}\sum_{\a}\overline{\av{\hat c\dg_{i\a}(t)\hat c_{j\a}(t')}}_{\ro_0},\\
% \label{eqA:GLesser}
% &&G^{>}_{ij}(t,t')=-\frac{i}{N}\sum_{\a}\overline{\av{\hat c_{j\a}(t')\hat c\dg_{i\a}(t)}}_{\ro_0}.
% \label{eqA:GGreater}
% \end{eqnarray}
% \label{eqA:GLesserGreater}
% \end{subequations}
% where $\overline{\av{\hat X}}_{\ro_0}=\overline{\tr{\hat X \ro_0}}$ denotes disorder and thermal averaging over an operator $\hat X$ with respect to the initial state, $\ro_0$. The operators in the Heisenberg picture are given by $\hat c_{ja}(t)=\hat U\dg(t)\hat c_{ja} \hat U(t)$, and $\hat U(t)=\cT e^{-i \int_0^t dt'\hat \cH(t')}$; $\cT$ is the time-ordering operator. The greater and the lesser Green's functions are related by $G^{>}_{ij}(t,t')=-[G^{<}_{ji}(t',t)]\dg$. \cBlue{Verify!}

Instead of evolving in time the  $G^R$ and $G^<$ functions, as appears in Eq.~\eqref{eq:KadanoffBaym}, we found it more convenient to evolve the retarded $G^R$, and Keldysh components $G^K$. The latter is defined as  
\Eq{
G_{ss'}^{K}(k;t,t')=G_{ss'}^{>}(k;t,t')+G_{ss'}^{<}(k;t,t').
\label{eqA:KeldyshGreenFunction}
}
% % \begin{subequations}
% % \begin{eqnarray}
% % &&G^{R}_{ij}(t,t')=\q(t-t')[G_{ij}^{>}-G_{ij}^{<}](t,t')\\
% % &&G^{K}_{ij}(t,t')=[G_{ij}^{>}+G_{ij}^{<}](t,t'),
% % \end{eqnarray}
% % \label{eqA:KeldyshGreenFunctions}
% % \end{subequations}
% where $\Q(t)$ is the Heaviside step function \cite{Rammer2007,Kamenev2011}. 
% The greater and lesser Green's functions can be recovered using the relation $G^{\gtrless}=\half\bS{G^K\pm (G^R- G^A)}$, where the advanced Green function is given by $G^A_{ij}(t,t')=\bS{G^R_{ji}(t',t)}\dg$. 
The Kadanoff-Baym equations for these two  components read
\begin{subequations}
\begin{eqnarray}
&&[i\dpa_t-H_0(t)] G^{R}=\dl(t-t')+\Si^R\circ G^R\label{eqA:KadanoffBaym1}\\
&&[i\dpa_t-H_0(t)] G^{K}=\Si^R\circ G^K+\Si^K\circ G^A\label{eqA:KadanoffBaym2}.
\end{eqnarray}
\label{eqA:KadanoffBaymModified}
\end{subequations}

The disorder-averaged self-energy for the chain of SYK dots with SYK-4 [Eq.~\eqref{eq:SYKHamiltonian}] and SYK-2 [Eq.~\eqref{eqA:HamiltonianSYK2}] interactions can be written in the self-consistent form. The diagrammatic structure of the self-energy to the leading order in $N$ is shown in Fig.~\ref{fig:SYKRelations} \cite{Maldacena2016}. As follows from this diagram, the greater and lesser components of the self-energy are given by 
\EqS{
\Si^{\gtrless}_{ss'}(x;t,t')=&K^2G^{\gtrless}_{ss'}(x;t,t')+\\
+&J^2 [G^{\gtrless}_{ss'}(x;t,t')]^2G_{s's}^{\lessgtr}(-x;t',t).
\label{eqA:SelfEnergy}
}
Here, $x$ is obtained from the Fourier transform of the momentum, $k$, index.
The retarded and Keldysh Green's functions $G^{R}$ and $G^{K}$ [appearing in Eq.~\eqref{eqA:KadanoffBaymModified}], are related to $G^{\gtrless}$ via 
$G^{R}_{ss'}(k;t,t')=\q(t-t')[G_{ss'}^{>}(k;t,t')-G_{ss'}^{<}(k;t,t')]$ and Eq.~\eqref{eqA:KeldyshGreenFunction}. In turn, the inverse relatons read
$G^{\gtrless}=\half\bS{G^K\pm (G^R- G^A)}$, where  $G^A_{ss'}(k;t,t')=G^R_{s's}(k;t',t)\dg$. The relations for the self energy are similar, with $G$ replaced by $\Si$.

\begin{figure}
  \centering
  \includegraphics[width=8.6cm]{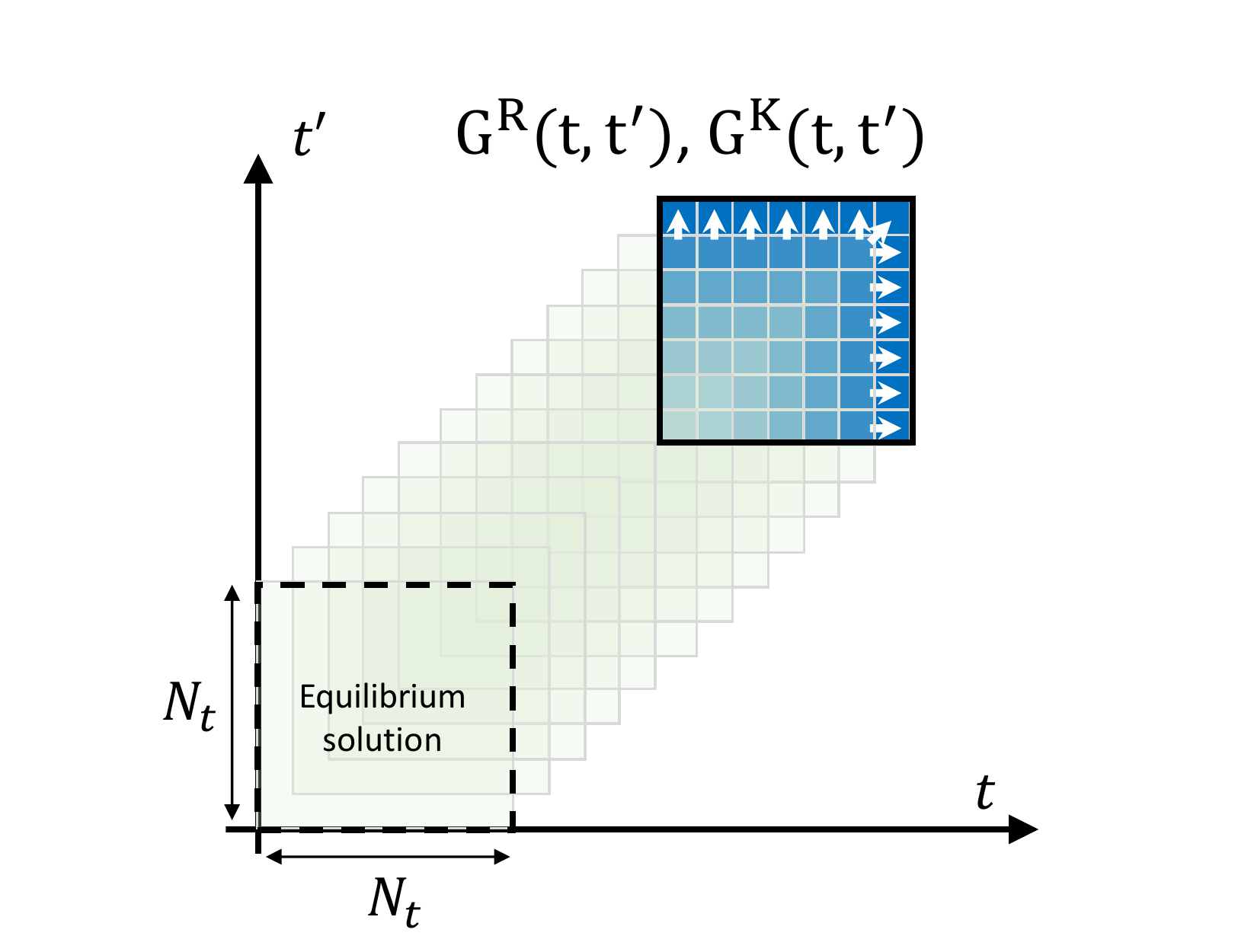}\\
  \caption{Time evolution of the retarded and Keldysh Green's functions according to the Kadanoff-Baym equations [Eqs.~\eqref{eqA:KadanoffBaymModified}]. The square in the bottom left corner represents the initial conditions for the Green's function, $G_{\rm eq}(t,t')$. At each step of the evolution a new row, column and a cell on the diagonal are added to the matrix, corresponding to $G(t_0+\dl t,t)$, $G(t,t_0+\dl t)$ and  $G(t_0+\dl t,t_0+\dl t)$.
%   b. Complex space structure of the integrand in the calculation of the topological invariant. At zero temperature the integral along the imaginary frequency is equivalent to counting the number of poles in the negative real half of the complex frequency space. At finite temperature, the integral along the real frequency is equivalent to counting all the poles on the real axis. Multiplying the function by the Fermi function (with poles on the imaginary axis at Matsubara values) weights each of the poles by its occupation probability.
\label{fig:TimeEvolution}}
\end{figure}

\subsection{Equilibrium solution}

Equations \eqref{eqA:KadanoffBaymModified} and \eqref{eqA:SelfEnergy} constitute a set of integro-differential equations which determine the time evolution of the Green's functions. Initial conditions for this time-evolution are set by the state $\ro_0$, corresponding to equilibrium with an inverse temperature $\be_0$ and Hamiltonian $\hat\cH(0)$. Due to invariance  to time-translations in equilibrium, the equilibrium Green's functions depend only on the time-difference $\D t=t-t'$. To find the equilibrium solution, we we evaluate Eq.~\eqref{eqA:KadanoffBaym1} for  $H_0(t)$ at $t=0$ and transform $\D t$ to the frequency space $\w$, giving rise to 
\Eq{
[\w-H_0(0)]G^R(\w)=1+\Si^R(\w)G^R(\w).
\label{eqA:EquilibriumEquation}
}
Furthermore, Eq.~\eqref{eqA:KadanoffBaym2} is trivially satisfied in equilibrium, due to the  fluctuation-dissipation theorem \cite{Chaikin1995,Rammer2007,Kamenev2011},
\Eq{
G^K_{ss'}(\w)=F(\w)[G^R_{ss'}(\w)-G^A_{ss'}(\w)],
\label{eqA:FluctuationDissipation}
}
where $F(\w)=\tanh(\be_0\w/2)$.
The equilibrium Green's function is obtained from the self-consistent solution of Eqs.~\eqref{eqA:SelfEnergy}, \eqref{eqA:EquilibriumEquation} and \eqref{eqA:FluctuationDissipation}.

% In this way, Eqs.~\eqref{eqA:KeldyshGreenFunctions}, \eqref{eqA:KadanoffBaym} and \eqref{eqA:SelfEnergy} in thermal equilibrium become a set of self-consistent equations in the frequency or time domain that can be solved numerically. In order to properly discretize the Green's functions,
% the grid size in the frequency domain ($d\w$) should be smaller than the minimal energy scale in the problem. At the same time, the frequency UV-cutoff ($\w_M$) should be larger than the maximal energy scale.

\subsection{Time evolution}
Having found an equilibrium solution, $G^R_{\rm eq}(k;t-t')$ and $G^K_{\rm eq}(k;t-t')$, we rearrange the vectors into matrices $[G_{\rm eq}^R]_{ss'}(k;t,t')$ and $[G^K_{\rm eq}]_{ss'}(k;t,t')$ of size $N_t\times N_t$ in the time domain and $2\times 2$ in the sublattice space, for a vector of crystal momenta of size $L$. In our simulations, we used $N_t=3000$ (smaller values of $N_t$ are used to vary $\dl \w$ and $\dl t$), and $L=50$. %, where $N_t=\lceil 2\w_M/d\w\rceil$. 
We used the equilibrium solution as the starting point of the simulation to propagate the Green's functions by one time step $\dl t$ in each iteration according to Eq.~\eqref{eqA:KadanoffBaymModified} \cite{Stan2009}, see Fig.~\ref{fig:TimeEvolution}.
In particular, given $G^R(t_0,t')$, we evolve $G^R$ according to:
\EqS{
&G^R(t_0+\dl t,t')=U_0(t_0)[G^R(t_0,t')-\frac{i}{2} \dl_{t,t'}]
-i\dl t I^R(t')
}
for $t'\le t_0$; $G^R(t,t_0+\dl t)=0$ for $t\le t_0$; and $G^R(t_0+\dl t,t_0+\dl t)=G^R(t_0,t_0)$. Here we defined $U_0(t)=e^{-i\dl t H_{0}(t)}$ and  $I^R(t')=\int \Si^R(t_0,s) G^R(s,t') ds$. 
% Similarly, we evolve $G^K$ using
% \EqS{
% &G^K(t_0+\dl t,t')=U_0(t)G^K(t_0,t')-i\dl t I^K(t').
% }
% for $t'\le t_0$; $G^K(t,t_0+dl t)=-[G^K(t_0+dt,t)]\dg$ for $t\le t_0$; and $G^K(t_0+dt,t_0+dt)=U_0(t_0) G^K(t_0,t_0)U_0\dg(t_0)-idt(I^K(t_0)+[I^K(t_0)]\dg)/2$.

To optimize the efficiency of the simulation, we keep the overall size of the matrices constants. Therefore, for each new element of the Green's function calculated in the future, we erase one element in the past. %Therefore, the memory of the system is restricted to the time-interval $T_\yM\sim N_t dt$. 
Such a truncation of the Green's function fixes the required memory of the simulation and significantly reduces the computational resources used.

%%%%%%%%%%%%%%%%%%%%%%%%%%%%%%%%%%%%%%%%%%%%%%%%
%%%%%%%%%%%%%%%%%%%%%%%%%%%%%%%%%%%%%%%%%%%%%%%%
%%%%%%%%%%%%%%%%%%%%%%Bibliography%%%%%%%%%%%%%%
%%%%%%%%%%%%%%%%%%%%%%%%%%%%%%%%%%%%%%%%%%%%%%%%
%%%%%%%%%%%%%%%%%%%%%%%%%%%%%%%%%%%%%%%%%%%%%%%%
%merlin.mbs apsrev4-1.bst 2010-07-25 4.21a (PWD, AO, DPC) hacked
%Control: key (0)
%Control: author (0) dotless jnrlst
%Control: editor formatted (1) identically to author
%Control: production of article title (0) allowed
%Control: page (1) range
%Control: year (0) verbatim
%Control: production of eprint (0) enabled
%

%%%%%%%%%%%%%%%%%%%%%%%%%%%%%%%%%%%%%%%%%%%%%%%%%%%
%%%%%%%%%%%%%%%%%%%%%End bibliography%%%%%%%%%%%%%%
%%%%%%%%%%%%%%%%%%%%%%%%%%%%%%%%%%%%%%%%%%%%%%%%%%%

\end{document}